\documentclass[twocolumn,amsmath,amssymb,prb]{revtex4}

\usepackage{graphicx}
\usepackage{dcolumn}
\usepackage{bm}

\begin{document}

\preprint{PRL}

\title{Two-photon- photoluminescence excitation spectroscopy of single quantum-dots}

\author{Y. Benny}
\email{byael@tx.technion.ac.il}
\affiliation{The Physics Department
and the Solid State Institute, Technion -- Israel institute of
technology, Haifa, 32000, Israel.}
\author{Y. Kodriano}
\affiliation{The Physics Department and the Solid State Institute,
Technion -- Israel institute of technology, Haifa, 32000, Israel.}
\author{E. Poem}
\affiliation{The Physics Department and the Solid State Institute,
Technion -- Israel institute of technology, Haifa, 32000, Israel.}
\author{S. Khatsevitch}
\affiliation{The Physics Department and the Solid State Institute,
Technion -- Israel institute of technology, Haifa, 32000, Israel.}
\author{P.~M.~Petroff}
\affiliation{Materials Department, University of California Santa
Barbara, CA, 93106, USA}
\author{D. Gershoni}
\affiliation{The Physics Department and the Solid State Institute,
Technion -- Israel institute of technology, Haifa, 32000, Israel.}

\date{\today}

\begin{abstract}
We present experimental and theoretical study of single
semiconductor quantum dots excited by two non-degenerate, resonantly
tuned variably polarized lasers. The first laser is tuned to
excitonic resonances. Depending on its polarization it
photogenerates a coherent single exciton state. The second laser is
tuned to biexciton resonances. By scanning the energy of the second
laser for various polarizations of the two lasers, while monitoring
the emission from the biexciton and exciton spectral lines, we map
the biexciton photoluminescence excitation spectra. The resonances
rich spectra of the second photon absorption are analyzed and fully
understood in terms of a many carrier theoretical model which takes
into account the direct and exchange Coulomb interactions between
the quantum confined carriers.
\end{abstract}

\pacs{Valid PACS appear here}
\maketitle

\section{\label{sec:Introduction}Introduction}
Semiconductor quantum dots (QDs) confine charge carriers in three
spatial directions. This confinement results in discrete spectrum of
energy levels and energetically sharp optical transitions
between these levels~\cite{Ediger07,Poem07}. These ``atomic-like" features, together with their compatibility with
modern semiconductor based microelectronics and optoelectronics,
make QDs promising building block devices for future technologies
involving single-photon emitters~\cite{Michler00} and quantum information
processing (QIP)~\cite{Zanardi98,Imamoglu99,Press10}.
In particular, QDs are considered to be an
excellent interface between photons, whose polarization state may
carry quantum information from one site to another, and confined
carriers' spins, whose states can be coherently manipulated
locally~\cite{Kosaka09,Benny10}. For these reasons, it is very important to study and to
understand in detail light-matter interactions in such
nanostructures. Deep understanding of these interactions is required
in order to implement protocols and schemes relevant to QIP~\cite{DiVincenzo95}, in
these man-made, technology compatible, `artificial atoms'.

In this work we present a comprehensive study of single, neutral
semiconductor QDs subject to excitation by two different
variably polarized resonant lasers. The first laser is tuned
to an excitonic resonance and generates a coherent single exciton
state, while the second laser is scanned through biexcitonic resonances.
Depending on the particular resonance and the direction of the light
polarization relative to the direction of the exciton spin, it
photogenerates a biexciton~\cite{Benny10}. The absorption is then monitored through
the emission intensities of various biexcitonic and excitonic
spectral lines.

The manuscript is organized as follows: Section \ref{sec:Theory} is
devoted to set the theoretical background which is required to
analyze the experimental data. In section \ref{sec:ExpSetup} we
describe the experimental methods and the measurements that we
performed. In section \ref{sec:Results} we present the experimental
results and analyze the data, using the theory outlined in section
\ref{sec:Theory}. The last part of this section provides a short
summary of the results.
\section{\label{sec:Theory}Theory}
We use a simple one-band model to describe the single-particle
wavefunctions of electrons in the conduction band and heavy-holes in
the valence band of a single QD. Since in these InAs/GaAs lattice
mismatch strain induced self-assembled QDs, the light-holes band is
energetically separated from the heavy-hole band by the strain and
the quantum size effect, light holes are not considered in our
model. The lateral extent of these QDs is typically about an order
of magnitude larger than their extent along the growth direction.
Therefore, for simplicity, our model considers only the two lateral
directions. The exact composition and strain distribution in these
QDs are not accurately known, therefore we use a very simple, two
dimensional parabolic potential model to describe the QD influence
on the carriers that it confines. This simple model is general
enough to describe the $C_{2\upsilon}$ symmetry of these
QDs~\cite{Singh10}, and it contains four parameters (see below)
which permit its adjustment to the experimental
observations~\cite{Poem07}. Two separated infinite elliptic
parabolic potentials are thus used, one for the electrons and one
for the heavy-holes~\cite{Warburton98}. The resulting envelope
wavefunctions or orbitals of the carriers, are therefore described
analytically by the 2D harmonic solutions:
\begin{equation}
\psi_{n_x,n_y}^{p}(x,y)=\frac{H_{n_x}(\frac{x}{l_p^x})H_{n_y}(\frac{y}{l_p^y})}{\sqrt{2^
{(n_x+n_y)}n_x!n_y!\pi l_p^x l_p^y}}e^{-\frac{1}{2}[(\frac{x}{l_p^x})^2+(\frac{y}{l_p^y})^2]}
\end{equation}
where p=e(h) stands for electron (heavy-hole) and $H_{n_{x(y)}}$ are
the Hermite polynomials of order $n_{x(y)}$.
\begin{equation}\label{eq:lchar}
l_p^{x(y)}=\sqrt{\frac{\hbar}{M^{*}_{\bot,p}\omega_p^{x(y)}}}
\end{equation}
is a characteristic length, which describes the extent of the
parabolic potential along the $x(y)$ direction and  Eq.\ \ref{eq:lchar} relates
this length to the charge carrier's in-plane effective mass
$M^{*}_{\bot,p}$ and the harmonic potential
inter-level energy separation $\hbar \omega_p^{x(y)}$ . We fit the four characteristic
lengthes $l_{e(h)}^{x(y)}$ to best describe the observed spectral
lines.

Equipped with the single carrier's eigenenergies and envelope wavefunctions we
proceed by calculating the many-carrier energies and states using
configuration interaction (CI)
model~\cite{Beranco95,Dekel0061,Poem07}. A detailed description of
the model, which takes into account the direct and exchange Coulomb
interactions between any pair of carriers in the QD, is presented
elsewhere~\cite{Poem07}.

Previous studies dealt mainly with optical transitions to ground
excitonic states as a tool to describe polarization sensitive photoluminescence (PL)
experiments~\cite{Poem07,Ediger07}. Here, motivated by our progress
in performing resonant PL excitation (PLE) spectroscopy, using one
and two laser sources, we use the same model to consider transitions
from various other levels. First, we consider transitions which
result from the resonant absorption of one photon. Then, we add a
second photon, resonantly tuned to the resonances of the optically
excited QD. Since, as we explain below, the situation in this case
is much richer than for PL only, we have to modify the notation for
describing the QD many-carrier states. We use the following
notation; A single carrier state is described by its envelope
wavefunction or orbital mode (O=1,2,...,6), where the number
represents the energy order of the level so that O=1 represents the
ground state. O is followed by the type of carrier, electron (e) or
heavy-hole (h) and a superscript which describes the occupation of
the single carrier state. The superscript can be either 1 (open
shell) or 2 (closed shell), subject to the Pauli exclusion principle
(non occupied states are not described). All the occupied states of
carriers of same type are then marked by subscripts which describe
the mutual spin configuration ($\sigma$) of these states.

\subsection{\label{sec:ExcitonCharac}Characterization of excitonic resonances}
\begin{figure}
\includegraphics[width=0.48\textwidth]{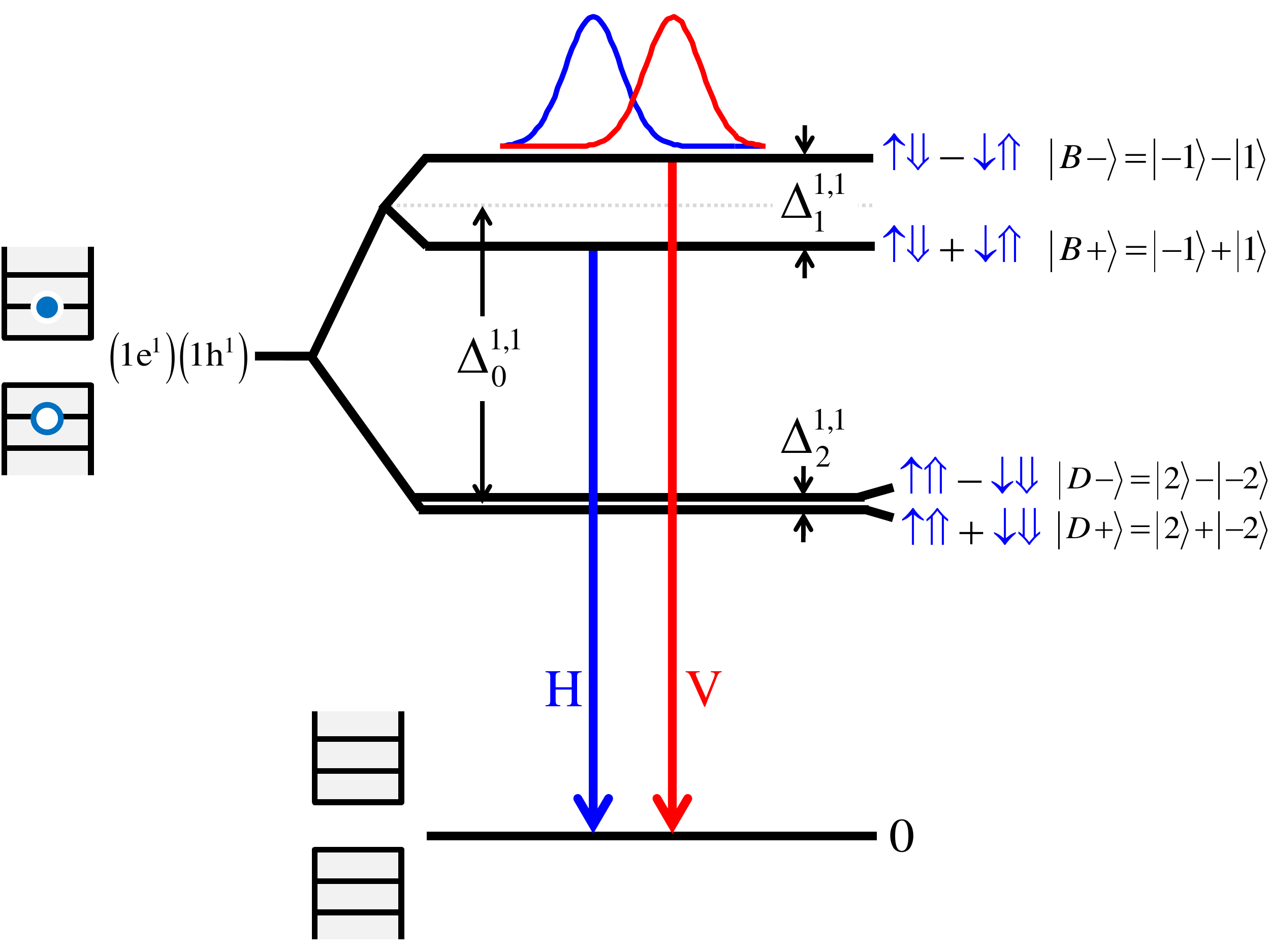}
\caption{\label{fig:X0} Schematic description of the energy levels
of the $\rm(1e^1)(1h^1)$ exciton and the allowed optical transitions
from its states to the vacuum. The major parts of the spin
wavefunctions are described to the right of each level.
$\uparrow$($\Downarrow$) represents an electron (hole) with spin up
(down) and a blue arrow represents a carrier in its first level. A
blue (red) vertical arrow represents linearly polarized optical
transition along the major (H) [minor (V)] axis of the QD. The
bracketed numbers stand for the total spin projection of the
carriers along the QD growth direction.}
\end{figure}

The ground exciton state [$\rm X^0_{1,1}\equiv(1e^1)(1h^1)$] is a
two-carrier state, formed mainly by one electron and one heavy-hole
in their respective ground states. The exchange interaction between
the electron and the heavy-hole ~\cite{Bayer02,Takagahara00,Poem07}
is  described, using the method of invariants~\cite{Ivchenko}, by
the following spin Hamiltonian~\cite{Bayer02}:
\begin{equation}
\label{eq:EffHam}
H_{X^0_{1,1}}=\sum_{i=x,y,z}(a_{i}^{1,1}S_{i}J_{i}+b_{i}^{1,1}S_{i}J^{3}_{i})
\end{equation}
where $S_{i}$ ($J_{i}$) denotes the $i^{th}$ cartesian component of
the electron (hole) spin and $a_{i}^{1,1}$ and $b_{i}^{1,1}$ are
spin-spin coupling constants. The total spin projection on the
$i^{th}$ direction is thereby given by $F_{i}=S_{i}+J_{i}$. Here as
well, the interaction with light-holes is neglected so that in
$J_{i}$ only heavy-hole spins are considered~\cite{Bayer02}. In
matrix form, for the basis $|S_{z}\rangle\otimes |J_{z}\rangle$:
\begin{equation}
\begin{tabular}{llll}
$|-1/2,3/2\rangle$&$=$&$\downarrow^1\Uparrow^1$&, $F_{z}=1$\\
$|1/2,-3/2\rangle$&$=$&$\uparrow^1\Downarrow^1$&, $F_{z}=-1$\\
$|1/2,3/2\rangle$&$=$&$\uparrow^1\Uparrow^1$&, $F_{z}=2$\\
$|-1/2,-3/2\rangle$&$=$&$\downarrow^1\Downarrow^1$&, $F_{z}=-2$\\
\end{tabular}
\end{equation}
where $\uparrow^j$ ($\Downarrow^j$) indicates spin-up (-down)
electron (heavy-hole) in the orbital $j$, the Hamiltonian is given
by
\begin{equation}
H_{X^0_{1,1}}=\frac{1}{2}
\left(\begin{array}{r|cccc}
&1&-1&2&-2\\
\hline
1&\Delta_{0}^{1,1}&\Delta_{1}^{1,1}&0&0\\
-1&\Delta_{1}^{1,1}&\Delta_{0}^{1,1}&0&0\\
2&0&0&-\Delta_{0}^{1,1}&\Delta_{2}^{1,1}\\
-2&0&0&\Delta_{2}^{1,1}&-\Delta_{0}^{1,1}\\
\end{array}\right).
\end{equation}
where $\Delta_{0}^{1,1}=3(a_z^{1,1}+2.25b_z^{1,1})$,
$\Delta_{1}^{1,1}=1.5(b_x^{1,1}-b_y^{1,1})$ and
$\Delta_{2}^{1,1}=1.5(b_x^{1,1}+b_y^{1,1})$~\cite{Bayer02}.
${\Delta_{0,1,2}^{1,1}}$ are three constants, which fully
characterize the exchange interaction between the carriers in the
ground states~\cite{Ivchenko}. It is clearly seen that in a $C_{2\upsilon}$ symmetry the exchange
interaction completely removes the degeneracy between the four
possible various combinations of the electron-hole pair spin states~\cite{Singh10}.
The eigen-energies and the eigenstates are schematically described
in Fig.\ \ref{fig:X0}. The lowest energy state is the symmetric dark
exciton state which in our notation is described as follows: $\rm
|X^{0}_{1,1,D_{+}}\rangle
\equiv\frac{1}{\sqrt{2}}[(1e^1)_{1/2}(1h^1)_{3/2}+(1e^1)_{-1/2}(1h^1)_{-3/2}]$.
$\Delta_{2}^{1,1}$ above it lies the anti-symmetric dark exciton
state: $\rm |X^{0}_{1,1,D_{-}}\rangle
\equiv\frac{1}{\sqrt{2}}[(1e^1)_{1/2}(1h^1)_{3/2}-(1e^1)_{-1/2}(1h^1)_{-3/2}]$.
$\Delta_{2}^{1,1}$, is known to be quite small and believed to be
orbit independent~\cite{Ivchenko}. It was recently measured from the
temporal period of the coherent precession of the dark exciton spin
to be 1.4 $\rm\mu eV$~\cite{Poem210}.
The bright exciton eigenstates in which the electron and heavy-hole
spins are anti-aligned lie $\Delta_{0}^{1,1}$ above the dark
exciton states. The isotropic e-h exchange, $\Delta_{0}^{1,1}$,
was previously found to be about 300 $\rm\mu eV$, by magneto-optical measurements~\cite{Braitbart06}.

The symmetric and antisymmetric bright exciton states $\rm
|X^{0}_{1,1,B \pm}\rangle
\equiv\frac{1}{\sqrt{2}}[(1e^1)_{-1/2}(1h^1)_{3/2}\pm(1e^1)_{1/2}(1h^1)_{-3/2}]$,
are split by the anisotropic e-h exchange, $\Delta_{1}^{1,1}$.
The magnitude and sign of $\Delta_{1}^{1,1}=-34$ $\rm\mu eV$, is directly
measured by polarization sensitive PL spectroscopy.
Since $\Delta_{1}^{1,1} $ is negative, the
antisymmetric state ($\rm |X^{0}_{1,1,B-}\rangle$) is higher in
energy than the symmetric one ($\rm |X^{0}_{1,1,B+}\rangle$)~\cite{Takagahara00,Poem07}.

Conservation of angular momentum dictates that when a
$\downarrow\Uparrow$ ($\uparrow\Downarrow$) e-h pair radiatively
recombines, a right- (left-) hand circularly polarized photon is
emitted. It follows that radiative recombination from the symmetric
(antisymmetric) bright exciton state is linearly polarized H (V)
along the major (minor) in-plane axis of the QD~\cite{Ivchenko,
Gammon96}. We note that the symmetric and antisymmetric dark and
bright exciton states are by no means unique to the first single
carrier spatial levels (Oe=Oh=1). In fact, similar bright and dark
excitonic states are formed for any combination of Oe and Oh single
carrier states. In general, ${\Delta_{0,1,2}^{Oe,Oh}}$ depend on the
orbital mode of the carriers~\cite{Takagahara00,Poem07}.

When the laser is resonantly tuned into one of the excited bright
exciton states and its light is polarized correctly, the light is
absorbed and a single electron-hole pair is photogenerated. For
example, let us consider the states: $\rm|X_{1,2,B
\pm}^0\rangle\equiv1/\sqrt{2}[(1e^1)_{-1/2}(2h^1)_{3/2} \pm
(1e^1)_{1/2}(2h^1)_{-3/2}]$. These states are similar to the ground
bright states $\rm|X_{1,1,B \pm }^0\rangle$, albeit, here the hole
is in its second orbital mode ($\rm Oh=2$). Electron and hole pair
will be photogenerated in these levels, and then the hole will
rapidly relax non-radiatively (within $\sim$20 psec~\cite{Poem10,
Kodriano10}), by emitting phonons, to the ground level ($\rm
Oe=Oh=1$). This relaxation is faster than the radiative
recombination rate ($\sim$1 nsec).

Experimental identification of single photon or single exciton
transitions is conventionally done by polarization sensitive PL and
PLE spectroscopies. In PL, a QD is optically excited. The excitation
gives rise to light emission from various long-lived states which do
not relax to lower energy states within their radiative lifetime.
Polarization and intensity sensitive PL spectroscopies are in
particular useful for these identifications. For example, the bright
exciton typically gives rise to PL doublet composed of two
cross-linearly polarized components. These components are due to
recombination from each of its non-degenerate eigenstates.

The second orbital wavefunction ($p_H$) has one node along the major
symmetry axis of the QD. As a result $\Delta_{1}^{1,2}$ is positive
and therefore the symmetric eigenstate $\rm |X_{1,2,B+}^0\rangle$ is
higher in energy than the anti-symmetric
eigenstate~\cite{Takagahara00}. It thus follows that the V linearly
polarized transition to this exciton is lower in energy than the H
polarized one. When the $\rm |X_{1,2,B-}^0\rangle$ ($\rm
|X_{1,2,B+}^0\rangle $) state is excited, the hole rapidly relaxes
non-radiatively to its ground state, releasing its energy into
acoustical phonons. Since phonons do not interact with the carriers'
spin~\cite{Poem10,Benny10,Kodriano10}, the spin wavefunction's
symmetry remains the same, and the recombination occurs from the
$\rm |X_{1,1,B-}^0\rangle$ ($\rm |X_{1,1,B+}^0\rangle $) state.
Thus, polarization sensitive PLE spectroscopy can be efficiently
used to identify and sort various excitonic resonances.
\subsection{\label{sec:ExcitonCharac}Characterization of biexcitonic resonances}
The ground biexciton state is formed mainly by two spin paired
electrons and two spin paired heavy-holes in their respective ground
states. In our notation this state is described as follows: $\rm
|XX_{1,1,1,1}^0\rangle\equiv(1e^{2})(1h^{2})$. We note that spin paired
carriers can only form an antisymmetric spin singlet state and
therefore the $\sigma$ subscripts in this case is redundant and it
is omitted from the pair's state description. For unpaired carriers,
however, the situation is different. Two unpaired carriers can form
either a one antisymmetric singlet state, or three symmetric triplet
states. Therefore, in this case we do assign $\sigma$ subscripts for
describing the unpaired carriers' wavefunctions. For two unpaired
carriers the $\sigma$ can either be S, to indicate a singlet state,
or $\rm T_m$, to indicate a triplet state. Here m is the total spin
projection, 0 or $\pm 1$ (0 or $\pm 3$) of the pair of electrons
(heavy-holes), along the QD growth direction. A full description of
a state with two unpaired electrons and two unpaired holes has
therefore the form
\begin{equation}
\small
\rm|XX_{O_{e1},O_{e2},O_{h1},O_{h2},\sigma_e,\sigma_h}^0\rangle\equiv(O_{e1}e^{1}O_{e2}e^{1})_{\sigma_e}(O_{h1}h^{1}O_{h2}h^{1})_{\sigma_h}.
\end{equation}
We note that for a given set of 4 unpaired spatial coordinates, 16
different states with different spin configurations are possible.
These are naturally divided into the following 4 subgroups; One
state, similar in nature to the ground biexciton state, in which the
two electrons form a singlet (e-singlet) and the two heavy-holes
also form a singlet (h-singlet). Three states in which the electrons
form an e-singlet and the holes triplet (h-triplet), three in which
the holes form a h-singlet and the electrons e-triplet and nine in
which both the electrons and holes form triplets
(e-triplet-h-triplet). These four subgroups have different energies
due to the exchange interactions between carriers of same charge.
The lowest energy level includes the 9 e-triplet-h-triplet states,
the two intermediate groups include the 6 e-triplet-h-singlet and
e-singlet-h-triplet states and the highest energy one includes only
a single e-singlet-h-singlet state.

\begin{figure}
\includegraphics[width=0.48\textwidth]{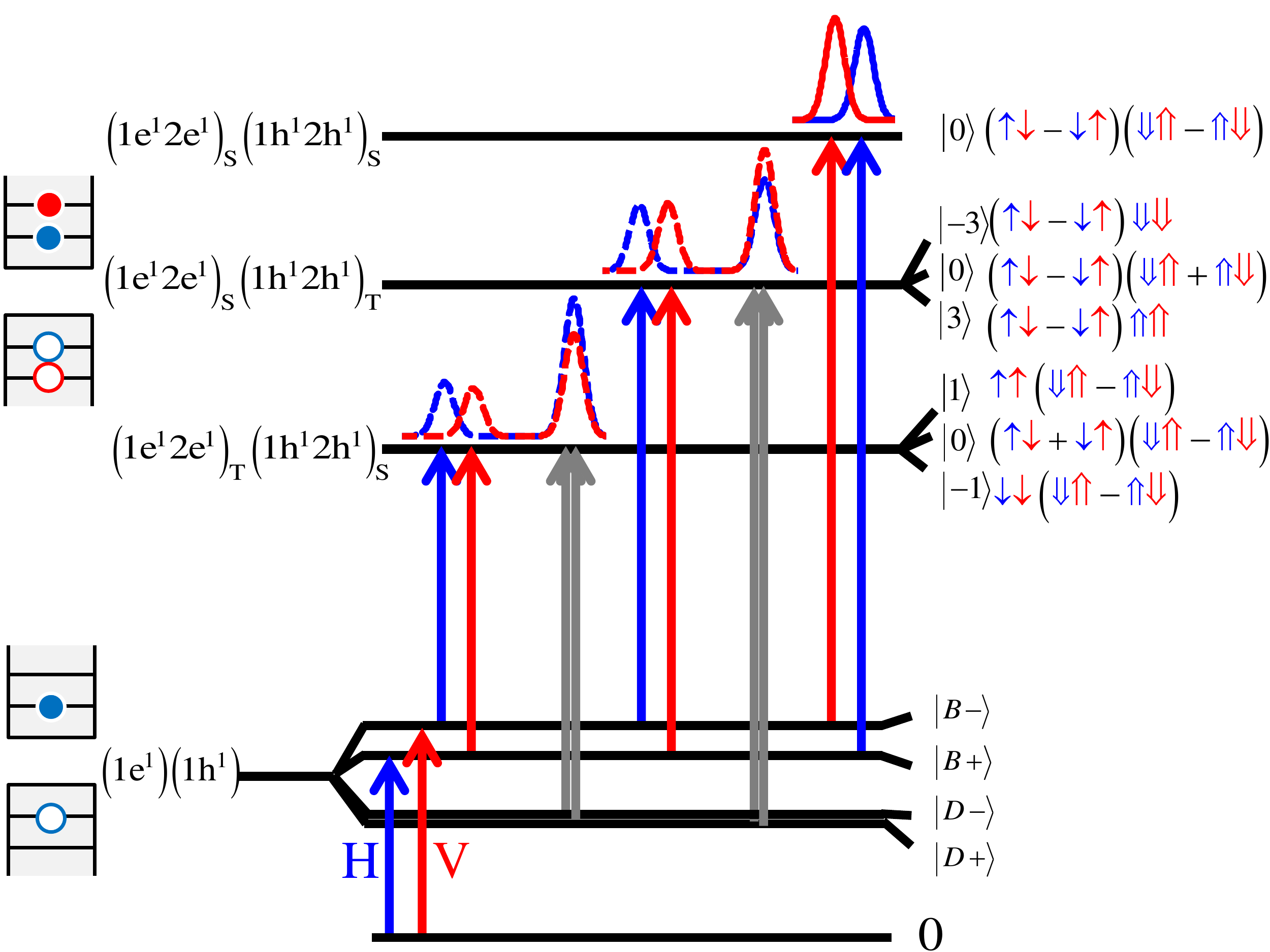}
\caption{\label{fig:XX0_ST_TS_SS} Schematic description of the
energy levels and spin wavefunctions of the configuration
$\rm(1e^12e^1)_{\sigma_e}(1h^12h^1)_{\sigma_h}$, for
$(\sigma_e,\sigma_h) =(T,S),(S,T),(S,S)$. The notations are as in
Fig.\ \ref{fig:X0}. Calculated two-lasers PLE spectra are presented by dash (solid) lines
for cross-(co-)linearly polarized exciton and biexciton transitions. Blue
(red) lines represent H (V) polarized biexcitonic transitions.}
\end{figure}

For simplicity, we begin by characterizing optical transitions in
which at least one type of carriers forms a singlet. In Fig.\
\ref{fig:XX0_ST_TS_SS} we schematically describe the energy levels
and the spin wavefunctions of the configuration
$\rm(1e^12e^1)_{\sigma_e}(1h^12h^1)_{\sigma_h}$, for the cases
$(\sigma_e,\sigma_h)=(T,S),(S,T)$, or $(S,S)$. The major parts of the
spin wavefunctions are described to the right of each level, where
$\uparrow$($\Downarrow$) represents an electron (hole) with spin up
(down) and a blue (red) arrow represents a carrier in its first
(second) level. The bracketed numbers stand for the total spin of
the configuration. H (V) polarized optical transitions are
represented by blue (red) vertical arrows.

We note that singlet-triplet and singlet-singlet biexcitonic
resonances may, in principle, occur also when the two carriers that
form the singlet reside in the same single carrier orbital mode (the
two carriers are paired). Naive intuitive considerations, which are
based on single-band models, predict that these transitions should
be weak, due to the small spatial overlap between the electron and
hole orbital modes which belong to different O
numbers~\cite{Bastard}. Transitions which involve orbital modes of
different symmetries should be forbidden in particular, since then,
their dipole moment vanishes. Nevertheless, these optically
forbidden transitions were previously observed in PLE spectroscopy
of quantum wells~\cite{Miller85} and QDs~\cite{Warming09,Siebert09}.

\begin{figure}
\includegraphics[width=0.48\textwidth]{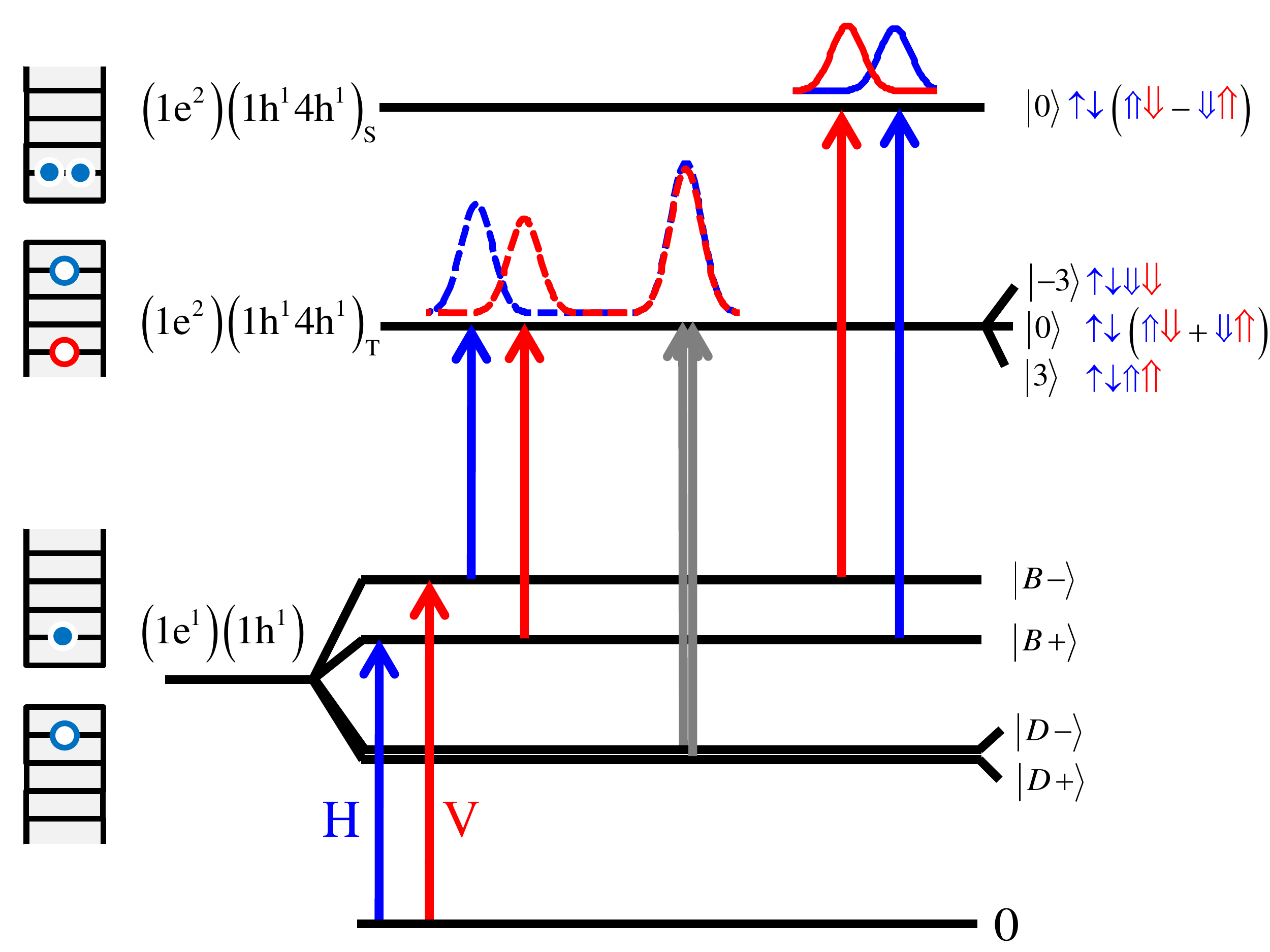}
\caption{\label{fig:XX0SD} Schematic description of the energy
levels and spin wavefunctions of the configuration
$\rm(1e^2)(1h^14h^1)$. The major parts of the spin wavefunctions are
described to the right of each level. The notations are as in Fig.\ \ref{fig:X0},
where $\uparrow$($\Downarrow$) represents an electron (hole) with
spin up (down) and a blue (red) arrow represents a carrier in its
first (excited) level. Calculated two-lasers PLE spectra are
presented by dash (solid) lines for cross- (co-) linearly polarized
excitonic and biexcitonic transitions. Blue (red) lines represent H
(V) polarized biexcitonic transitions.}
\end{figure}

In Fig.\ \ref{fig:XX0SD} we present an example for the case in which
the electrons are paired in their ground single carrier level while
the holes are not. One hole is in the Oh=1 s-like orbital and the
other is in the Oh=4, $\rm d_{HH}$-like orbital. Since the electrons
here are paired, they form a singlet, thus their total spin
vanishes. Therefore, the e-h exchange interaction is not expected to
remove the degeneracy between the holes triplet states. We find,
however, that this degeneracy is slightly removed due to
many-carrier mixing effects. Previous works attributed this effect
to anisotropic h-h exchange interactions~\cite{Ediger07,Warming09}, which our model does not contain.

Turning to Fig.\ \ref{fig:XX0_ST_TS_SS} again, we note that two
absorption resonances are expected from the bright exciton states
into an e-singlet-h-singlet state. These two transitions form a
typical cross-linearly polarized doublet, resembling the optical
transitions from the vacuum to the bright exciton states (see
Fig.\ \ref{fig:X0}). Four transitions are excepted from the exciton
states into the three e-singlet-h-triplet states and four similar
ones into the e-triplet-h-singlet states. Two of these four are
cross-linearly polarized transitions from the bright exciton states
into the state in which the two holes (or electrons) spins are
anti-parallel ($T_0$), and two cross-linearly polarized transitions
from each one of the dark exciton states into the corresponding
symmetric and anti-symmetric combinations of the parallel hole (or
electron) spin states $T_{\pm 3}$  ($T_{\pm 1}$) of the biexciton.
By inspecting the wavefunctions of the initial and final state of
each optical transition one immediately sees that the oscillator
strength of the optical transitions from the bright exciton states
is exactly half that of the transitions from the dark exciton
states. Moreover, since both the dark exciton and corresponding
biexciton pair states are nearly degenerate, these two transitions
form one unpolarized spectral line. Therefore, the total intensity
of this line is four times larger than that of the other two
transitions. The calculated spectra are presented in Fig.\
\ref{fig:XX0_ST_TS_SS} and Fig.\ \ref{fig:XX0SD}. In obtaining these
spectra, the calculated transition energies are convoluted with
normalized Gaussians of 50 $\rm\mu eV$ width, to take into account
the finite lifetime of the spin blockaded biexcitons. Transitions in
which the exciton and biexciton photons are co-(cross-) linearly
polarized are presented by solid- (dash-) lines, where blue- (red-)
lines represent H- (V-) polarized biexciton photons.

\begin{figure}
\includegraphics[width=0.48\textwidth]{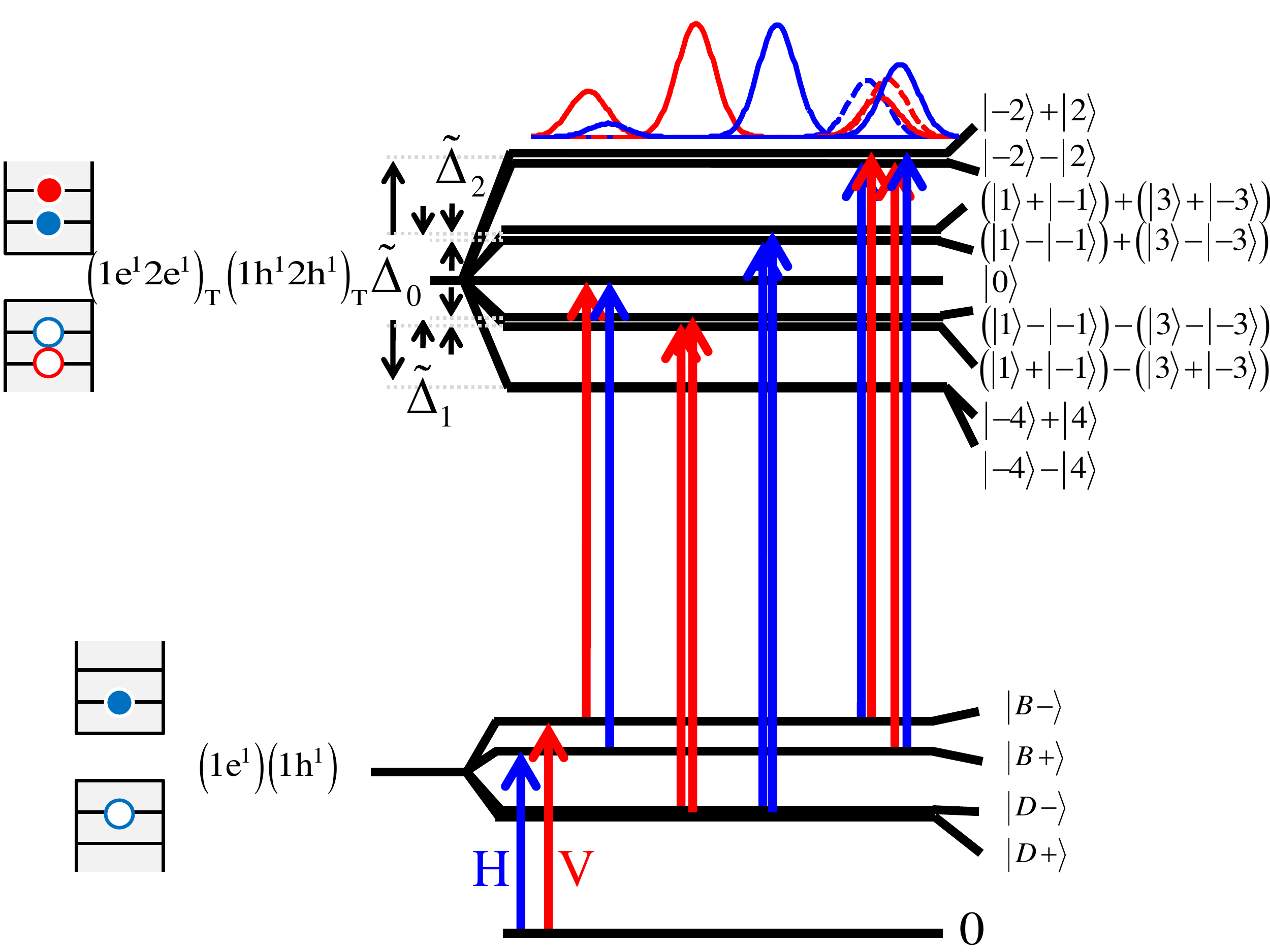}
\caption{\label{fig:XX016} Schematic description of the energy
levels of the $\rm(1e^12e^1)_{T}(1h^12h^1)_{T}$
biexciton and their optical transitions. The major parts of the spin
wavefunctions are described to the right of each level.
A blue (red) vertical arrow represents linearly polarized
optical transition along the major (H) [minor (V)] axis of the QD.}
\end{figure}

We now turn to discuss the optical transitions into
the e-triplet-h-triplet states. The electron-hole (e-h) exchange
interactions, which in our QDs are typically about an order of
magnitude smaller than same-carrier exchange interactions, remove
the degeneracy between the states within this subgroup. We
actually calculate the eigenenergies and eigenstates accurately
using a CI model~\cite{Dekel0061,Poem07}. However, for a more intuitive
discussion one can build an effective biexciton e-h exchange
Hamiltonian for the subspace of $\rm
(1e^12e^1)_{T_{e}}(1h^12h^1)_{T_{h}}$, using the single exciton
effective e-h exchange Hamiltonian of Eq.\ \ref{eq:EffHam}, such that an element is defined as
follows~\cite{Maialle07}:
\begin{equation}
\begin{tabular}{l}
$_{f}\langle J_{z}^2,J_{z}^1,S_{z}^2,S_{z}^1|H_{XX^{0}_{1,2,1,2}}|S_{z}^1,S_{z}^2,J_{z}^1,J_{z}^2\rangle_{i}$\\
$~~~~$=$~~$$_{f}\langle J_{z}^1,S_{z}^1|H_{X^{0}_{1,1}}|S_{z}^1,J_{z}^1\rangle_{i}$\\
$~~~~~~$+$_{f}\langle J_{z}^2,S_{z}^1|H_{X^{0}_{1,2}}|S_{z}^1,J_{z}^2\rangle_{i}$\\
$~~~~~~$+$_{f}\langle J_{z}^1,S_{z}^2|H_{X^{0}_{2,1}}|S_{z}^2,J_{z}^1\rangle_{i}$\\
$~~~~~~$+$_{f}\langle
J_{z}^2,S_{z}^2|H_{X^{0}_{2,2}}|S_{z}^2,J_{z}^2\rangle_{i}$,
\end{tabular}
\end{equation}
where $H_{X^{0}_{i,j}}$ is the single e-h pair spin Hamiltonian for
electron and hole in the orbital modes $i$ and $j$ respectively, and
the subscript $i$ ($f$) denotes the initial (final) spin state. We
change to a new basis in which the same-carrier exchange states are
diagonal. The weak e-h exchange interactions are then treated as
perturbations on these states. A similar approach was previously
used for describing charged excitons
(trions)~\cite{Kavokin03,Akimov05}. Taking only the subspace of the
e-triplet-h-triplet spin states, $|\sigma_{e}\rangle\otimes
|\sigma_{h}\rangle$:
\begin{equation}
\begin{tabular}{llll}
$|-1,3\rangle$&$=$&$\downarrow^1\downarrow^2\Uparrow^1\Uparrow^2$& $F_{z}=2$\\
$|-1,0\rangle$&$=$&$\downarrow^1\downarrow^2\frac{(\Downarrow^1\Uparrow^2+\Uparrow^1\Downarrow^2)}{\sqrt{2}}
$& $F_{z}=-1$\\
$|-1,-3\rangle$&$=$&$\downarrow^1\downarrow^2\Downarrow^1\Downarrow^2$&
$F_{z}=-4$\\
$|0,3\rangle$&$=$&$\frac{(\uparrow^1\downarrow^2+\downarrow^1\uparrow^2)}{\sqrt{2}}\Uparrow^1\Uparrow^2$&
$F_{z}=3$\\
$|0,0\rangle$&$=$&$\frac{(\uparrow^1\downarrow^2+\downarrow^1\uparrow^2)(\Downarrow^1\Uparrow^2+\Uparrow^1\Downarrow^2)}{2}
$& $F_{z}=0$\\ $|0,-3\rangle$&$=$&$\frac{(\uparrow^1\downarrow^2+\downarrow^1\uparrow^2)}{\sqrt{2}}\Downarrow^1\Downarrow^2$& $F_{z}=-3$\\
$|1,3\rangle$&$=$&$\uparrow^1\uparrow^2\Uparrow^1\Uparrow^2$& $F_{z}=4$\\
$|1,0\rangle$&$=$&$\uparrow^1\uparrow^2\frac{(\Downarrow^1\Uparrow^2+\Uparrow^1\Downarrow^2)}{\sqrt{2}}$& $F_{z}=1$\\
$|1,-3\rangle$&$=$&$\uparrow^1\uparrow^2\Downarrow^1\Downarrow^2$& $F_{z}=-2$\\
\end{tabular}
\end{equation}
we obtain the following matrix (neglecting many-body mixing corrections)

\begin{equation}
\label{eq:HamiltonianTT}
 H_{XX^0_{TT}}=\frac{1}{2}
 \hbox{\scriptsize$
\left(\begin{array}{r|ccccccccc}
&2&-1&-4&3&0&-3&4&1&-2\\
\hline
2&\widetilde{\Delta}_{0}&0&0&0&\widetilde{\Delta}_{1}&0&0&0&0\\
-1&0&0&0&\widetilde{\Delta}_{2}&0&\widetilde{\Delta}_{1}&0&0&0\\
-4&0&0&-\widetilde{\Delta}_{0}&0&\widetilde{\Delta}_{2}&0&0&0&0\\
3&0&\widetilde{\Delta}_{2}&0&0&0&0&0&\widetilde{\Delta}_{1}&0\\
0&\widetilde{\Delta}_{1}&0&\widetilde{\Delta}_{2}&0&0&0&\widetilde{\Delta}_{2}&0&\widetilde{\Delta}_{1}\\
-3&0&\widetilde{\Delta}_{1}&0&0&0&0&0&\widetilde{\Delta}_{2}&0\\
4&0&0&0&0&\widetilde{\Delta}_{2}&0&-\widetilde{\Delta}_{0}&0&0\\
1&0&0&0&\widetilde{\Delta}_{1}&0&\widetilde{\Delta}_{2}&0&0&0\\
-2&0&0&0&0&\widetilde{\Delta}_{1}&0&0&0&\widetilde{\Delta}_{0}\\
\end{array}\right)$}
\end{equation}
where
$\rm \widetilde{\Delta}_{0}=\frac{\Delta_{0}^{Oe_1,Oh_1}+\Delta_{0}^{Oe_1,Oh_2}+\Delta_{0}^{Oe_2,Oh_1}+\Delta_{0}^{Oe_2,Oh_2}}{4}$
and $\rm \widetilde{\Delta}_{1,2}=\frac{\Delta_{1,2}^{Oe_1,Oh_1}+\Delta_{1,2}^{Oe_1,Oh_2}+\Delta_{1,2}^{Oe_2,Oh_1}+\Delta_{1,2}^{Oe_2,Oh_2}}{8}$.
In Table \ref{tb:TT} we present the nine eigenenergies and
eigenfunctions of the effective Hamiltonian $H_{XX^0_{TT}}$.
These eigenenergies and spin wavefunctions are also presented in
Fig.\ \ref{fig:XX016}. The allowed optical transitions between the
ground exciton states to these biexciton states, together with their
polarization selection rules, are presented as well. We note that
since a photon can carry angular momentum of $\pm 1$ only, biexciton
resonances of total spin 3 and 1 can be reached optically only from
the ground dark exciton states. Similarly, biexciton resonances of
total spin 0 and 2 can be reached optically from the bright exciton
states only. Biexciton states with total spin 4 cannot be reached
optically.

\begin{table}
\caption{\label{tb:TT} Calculated eigen-energies and their
respective spin configurations for the e-triplet-h-triplet states.
The base states are expressed by their total spin projection $|F_z
\rangle$. }
\begin{ruledtabular}
\begin{tabular}{ll}
Energy&Configuration\\
\hline
$-\widetilde{\Delta}_0$&$|4\rangle-|-4\rangle$\\
$-(\widetilde{\Delta}_1+\widetilde{\Delta}_2)/2$&$(|1\rangle+|-1\rangle)-(|3\rangle+|-3\rangle)$\\
$-(\widetilde{\Delta}_1-\widetilde{\Delta}_2)/2$&$(|1\rangle-|-1\rangle)-(|3\rangle-|-3\rangle)$\\
$(\widetilde{\Delta}_1-\widetilde{\Delta}_2)/2$&$(|1\rangle-|-1\rangle)+(|3\rangle-|-3\rangle)$\\
$(\widetilde{\Delta}_1+\widetilde{\Delta}_2)/2$&$(|1\rangle+|-1\rangle)+(|3\rangle+|-3\rangle)$\\
$\widetilde{\Delta}_0$&$-(|2\rangle+|-2\rangle)$\\
$R_i$\footnote{$R_i$ is the $i^{th}$ root of the equation
$2R_i^3-(2\widetilde{\Delta}_0^2+2\widetilde{\Delta}_1^2+\widetilde{\Delta}_2^2)R_i+\widetilde{\Delta}_0(\widetilde{\Delta}_2^2-\widetilde{\Delta}_1^2)=0$.
The expressions in square brackets are obtained for the case
$\widetilde{\Delta}_2\ll\widetilde{\Delta}_1\ll\widetilde{\Delta}_0$. $R_i$ is given to order $(\widetilde{\Delta}_1/\widetilde{\Delta}_0)^2$.}, $i=1,2,3$&\\
$R_1\rightarrow [-\widetilde{\Delta}_0]$&$\rightarrow[|4\rangle+|-4\rangle]$\\
$R_2\rightarrow[\widetilde{\Delta}_0-\sqrt{\widetilde{\Delta}_0^2+2\widetilde{\Delta}_1^2}/2]$&$\rightarrow [|0\rangle]$\\
$R_3\rightarrow[\widetilde{\Delta}_0+\sqrt{\widetilde{\Delta}_0^2+2\widetilde{\Delta}_1^2}/2]$&$\rightarrow [|2\rangle+|-2\rangle]$\\
\end{tabular}
\end{ruledtabular}
\end{table}

\subsection{\label{sec:TheoManyBody}Many-carrier mixing effects}
The above discussion assumes that, to first order, the interactions
between the carriers are much smaller in comparison with the
single-carrier level separations. Therefore, we
safely ignore contributions to the biexciton eigenstates which
results from mixing with other configurations outside the subspace
considered. Our model, however, does include these
contributions~\cite{Dekel0061,Poem07} and as we show below, in some
cases, specifically when otherwise the transitions are forbidden,
mixing with other configurations are directly observed in the
experimental data.

Our model includes six orbital modes for each carrier. The
many-carrier eigenstates are obtained by the diagonalization of the
many-body Hamiltonian, which is constructed from all the possible
configurations of the confined carriers in a system of six bound
levels. Thus, a many-carrier eigenstate always contains
contributions from different combinations of single carriers'
orbital modes.

An example for transitions in which these contributions become
important are the optical transitions from the
$\rm(1e^12e^1)_{\sigma_e}(1h^2)$ biexciton to the first excited
exciton state where the leading contribution comes from the
configuration $\rm(1e^1)(2h^1)$. Our model calculation resulted in
optical transitions between these states, as indeed we found
experimentally (see below). In Fig.\ \ref{fig:XX0TeShSeTh} we
describe the energy level structure of the $\rm(1e^2)(1h^12h^1)_T$
and $\rm(1e^12e^1)_T(1h^2)$ biexcitons. Since these excited
biexciton states are spin blockaded for thermal relaxation, they
decay radiatively by recombination of a ground state e-h pair. The
optical transitions originated from their decay are also described
in Fig.\ \ref{fig:XX0TeShSeTh}. If one neglects mixing, it follows
that the $\rm(1e^2)(1h^12h^1)_T$ biexcitons decay into excited
$\rm(1e^1)(2h^1)$ excitons and the $\rm(1e^12e^1)_T(1h^2)$
biexcitons decay into $\rm(2e^1)(1h^1)$ excitons. These two excited
excitons are, however, highly mixed due to the Coulomb interaction
between the electron and the hole. Roughly speaking, our model shows
that each biexciton group decays into both excited exciton states,
resulting in four sets of three spectral lines. Each group of three
spectral lines resembles the sets described in Fig.\
\ref{fig:XX0_ST_TS_SS} and Fig.\ \ref{fig:XX0SD}. It contains two
lower energy cross-linearly polarized lines and one, four fold
stronger, unpolarized line. The calculated PL spectra which result
from these transitions are presented in Fig.\ \ref{fig:XX0TeShSeTh}.
The spectral width of the lines which results from emission to the
lower energy excited exciton states [mainly $\rm(1e^1)(2h^1)$] are
obtained by convoluting the calculated transitions with Gaussians of
$\sim 50$ $\rm\mu eV$, accounting for the finite lifetime of the
excited hole states. Similarly, the spectral width of the lines
which result from emission to the higher energy excited exciton
states [mainly $\rm(2e^1)(1h^1)$], should have been obtained by
convolution with Gaussians of $\sim 1$ meV (not visible, because the
convolution results in a nearly uniform, unpolarized background on
the relevant energy scale), due to the much shorter lifetime of the
excited electron states. The difference between the two cases is due
to the difference between the relaxation rates of the hole and the
electron. While a hole in the second orbital state relaxes
non-radiatively to the first orbital within $\sim 20$ psec by
emitting acoustical phonons~\cite{Poem10, Kodriano10}, the electron
does so within less than 1 psec, by coupling to optical
phonons~\cite{Hameau99}. The decay of the electrons is so rapid because the energy of
optical phonons in the wetting layer, nearly resonate with the
energy separation between the two electronic orbitals ($\sim$29
meV~\cite{Benny10}).
\begin{figure}
\includegraphics[width=0.48\textwidth]{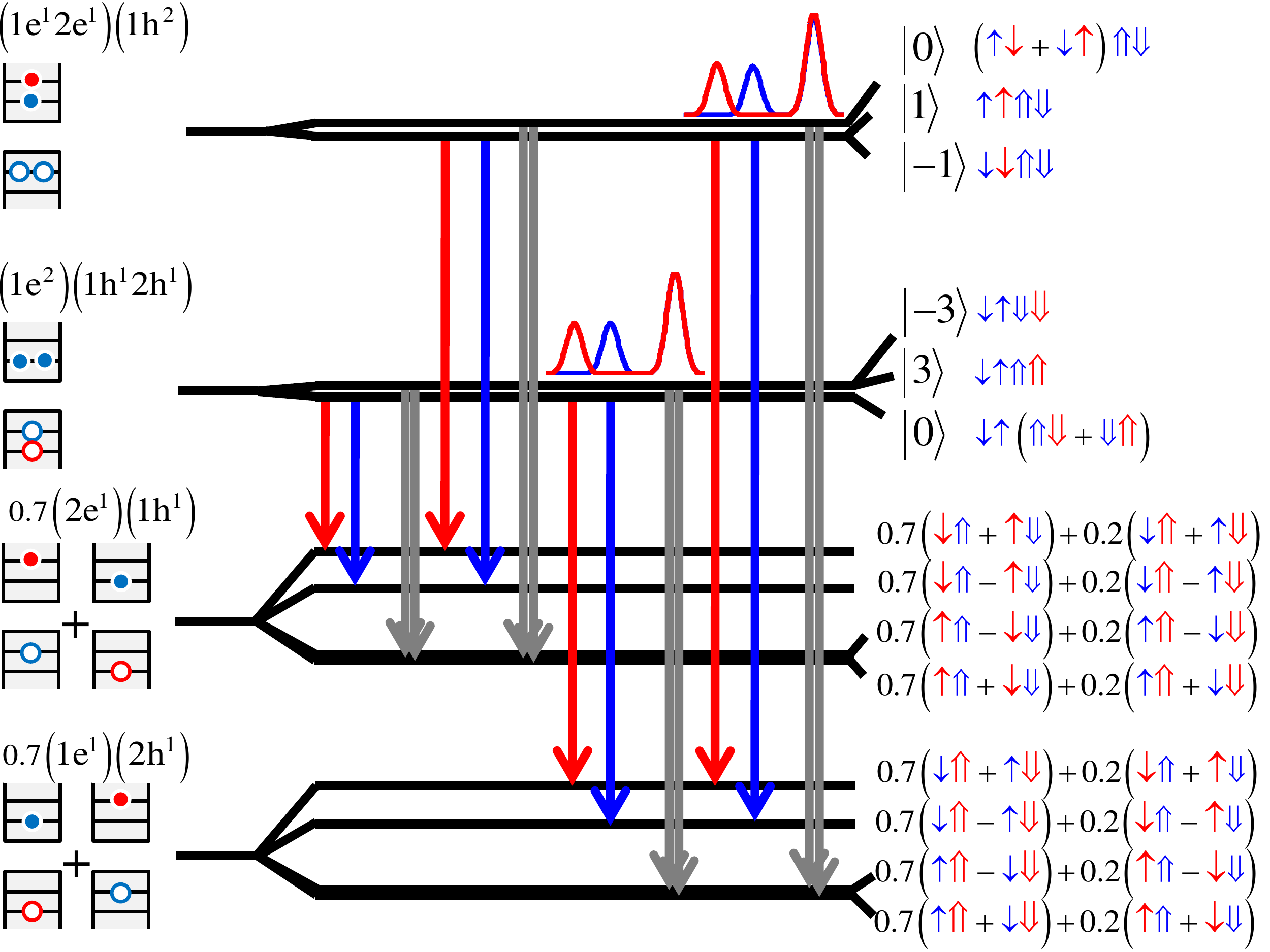}
\caption{\label{fig:XX0TeShSeTh} Schematic description of the energy
levels of $\rm(1e^2)(1h^12h^1)_T$ and $\rm(1e^12e^1)_T(1h^2)$
biexcitons and the excited exciton states $\rm(1e^1)(2h^1)$ and
$\rm(2e^1)(1h^1)$. The notations used here are the same as in Fig.\
\ref{fig:XX0SD}. The curves describe calculated PL spectra of the
transitions to the $\rm(1e^1)(2h^1)$ excited excitons. The
calculated transitions to the $\rm(2e^1)(1h^1)$ excited excitons are
not visible due to the short lifetime of the final state.}
\end{figure}

\section{\label{sec:ExpSetup}The experimental setup}

The sample used in this work was grown by molecular-beam epitaxy on
a (001) oriented GaAs substrate. One layer of strain-induced $\rm
In_{x}Ga_{1-x}As$ QDs was deposited in the center of a one
wavelength microcavity formed by two unequal stacks of alternating
quarter wavelength layers of AlAs and GaAs, respectively. The height
and composition of the QDs were controlled by partially covering the
InAs QDs with a 3 nm layer of GaAs and subsequent growth
interruption. To improve photon collection efficiency, the
microcavity was designed to have a cavity mode, which matches the QD
emission due to ground state e-h pair recombinations. During the
growth of the QD layer the sample was not rotated, resulting in a
gradient in the density of the formed QDs. The estimated QD density
in the sample areas that were measured is $10^8$ $\rm cm^{-2}$; however,
the density of QDs that emit in resonance with the microcavity mode
is more than two orders of magnitude lower~\cite{Ramon06}. Thus,
single QDs separated apart by few tens of micrometers were easily
located by scanning the sample surface during PL measurements.
Strong anti-bunching in intensity auto-correlation measurements
were then used to verify that the isolated QDs are single ones and
that they form single photon sources.

\begin{figure}
\includegraphics[width=0.48\textwidth]{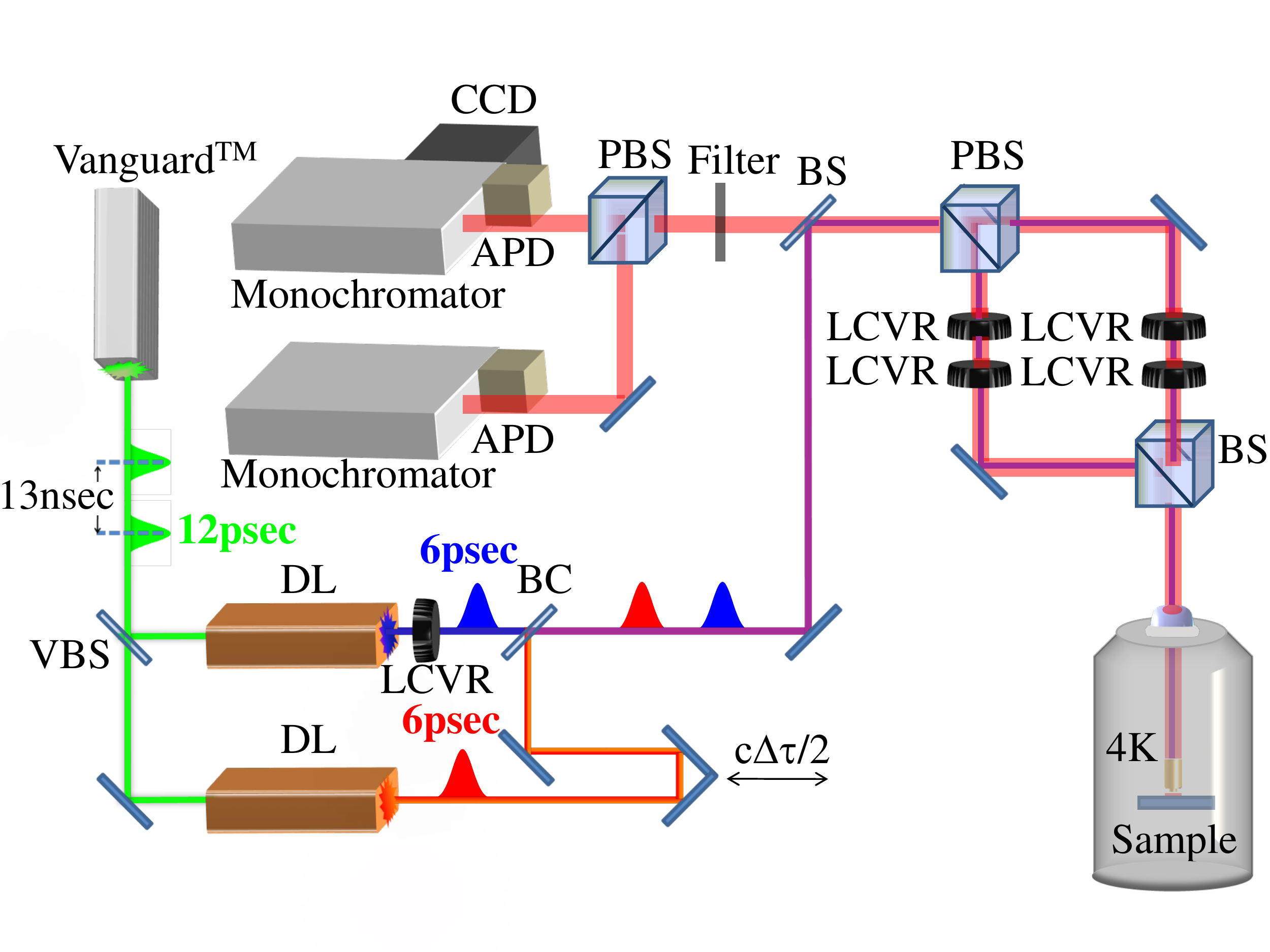}
\caption{\label{fig:ExpSetup} Schematic description of the
experimental setup. The two dye laser pulses can be delayed one with
respect to the other by a retroreflector mounted on a
computer-controlled translation stage. For the cw experiments
Ti:sapphire lasers were used. (P)BS stands for (polarizing) beam
splitter; VBS for a variable beam splitter; BC for beam combiner;
and LCVR for a liquid crystal variable retarder.}
\end{figure}

The experimental setup that we used for the optical measurements is described in Fig.\
\ref{fig:ExpSetup}. The sample was placed inside a sealed metal tube
immersed in liquid Helium, maintaining temperature of 4.2K. A
$\times$60 microscope objective with numerical aperture of 0.85 was
placed above the sample and used to focus the light beams on the
sample surface and to collect the emitted PL. The majority of this
work was performed with cw excitation. We used one tunable
Ti:sapphire laser to scan the energy. A second Ti:sapphire laser was
used for the two-photon excitation experiments. We performed also
measurements with pulse excitation. In these measurements we used
two dye lasers, synchronously pumped by the same frequency-doubled
Nd:YVO$\rm_4$ (Spectra Physics-Vanguard$\rm^{TM}$) laser for
generating the resonantly tuned optical pulses, as described in the
figure. The repetition rate of the setup was 76 MHz, corresponding to
a pulse separation of about 13 nsec. The duration of the laser
pulses were about 6 psec and their spectral widths about 200 $\rm\mu
eV$. The delay between the pulses was controlled by a retroreflector
on a translation stage.

The lasers emission energy could have been continuously tuned using
coordinated rotations of two plate birefringent filters and a thin
etalon. The polarizations of the pulses were independently adjusted
using a polarized beam splitter (PBS) and two pairs of computer
controlled liquid crystal variable retarders (LCVRs). The
polarization of the emitted PL was analyzed by the same LCVRs and
PBS.  The PL was spectrally analyzed by 1-meter monochromator and
detected by either a silicon avalanche photodetector or by a cooled
CCD camera.

In polarized PLE spectroscopy, one monitors the polarized emission from an
identified PL line while varying the energy and polarization of
the exciting light source. From the variations in the intensity of
the emitted PL, one can readily identify many carrier resonances in
which the light is preferentially absorbed. Increased absorption,
which results in increased emission intensity of a specific PL line,
and its polarization sensitivity are then used to unambiguously
identify the many-carrier state which forms a specific absorption
resonance~\cite{Finley01,Ware05}.
\section{\label{sec:Results}Results}
\begin{figure}
\includegraphics[width=0.48\textwidth]{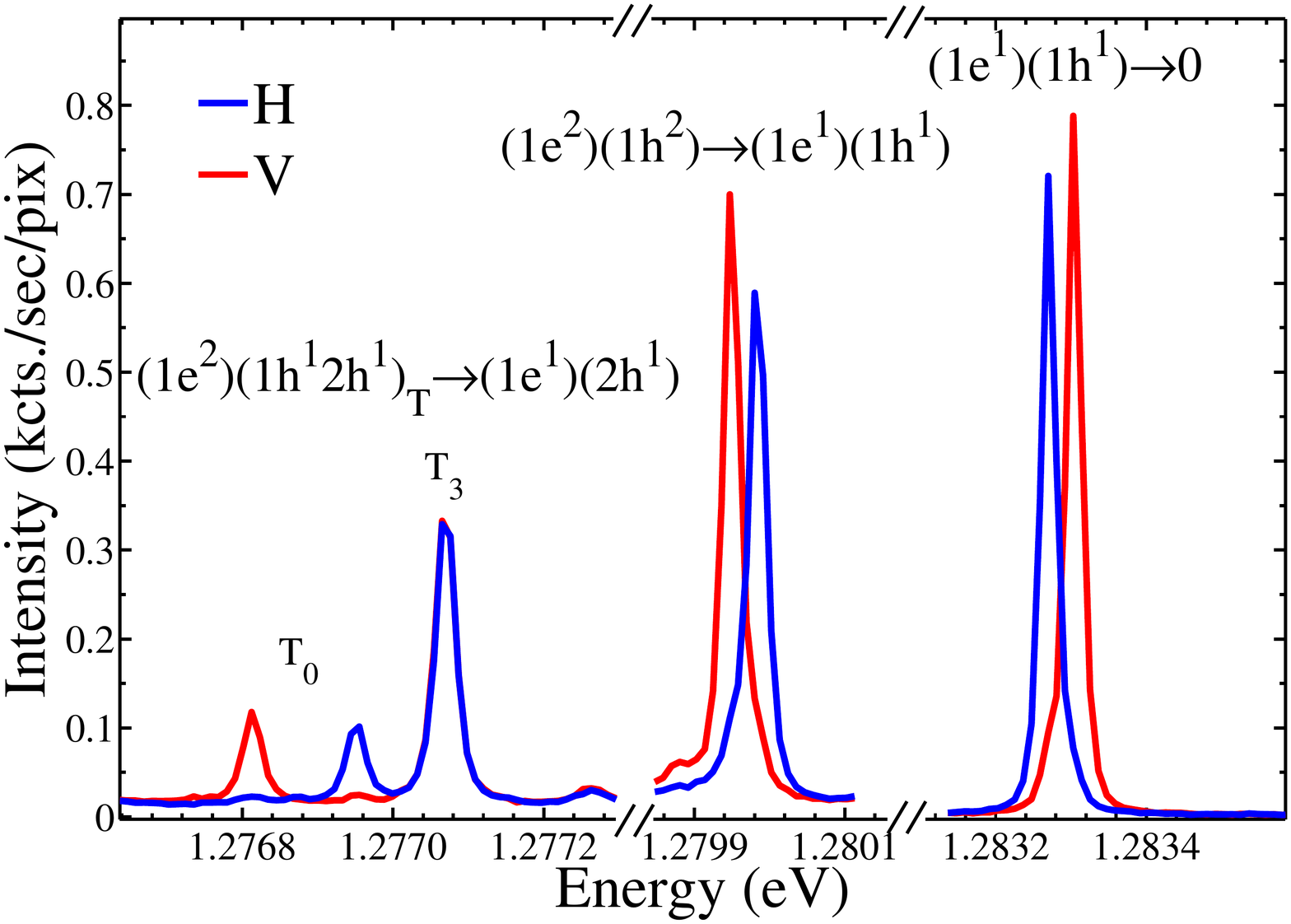}
\caption{\label{fig:PL} Linear polarization sensitive PL spectra,
showing the neutral exciton and biexciton lines of a single QD
excited by a 501 nm cw laser. The spectral transitions are
identified in the figure.}
\end{figure}

In Fig.\ \ref{fig:PL} we present polarization sensitive PL spectrum
of a single QD in resonance with the microcavity mode. The PL was
obtained by exciting the QD with a 501 nm cw Ar+ laser. We found
that at this excitation energy the QD is on average
neutral~\cite{Kodriano10}. The excitation intensity was roughly $\rm
1 W/cm^2$ aiming at obtaining equal emission intensity from the
exciton and biexciton lines~\cite{Dekel00}. The spectral neutral
excitonic and biexcitonic lines, which are relevant for this study
are identified above the spectral features in the figure. We note
that in addition to the ground bright exciton ($\rm |X^{0}_{1,1,B
\pm}\rangle$ to vacuum) and ground biexciton [$\rm(1e^2)(1h^2)$ to $|X^{0}_{1,1,B
\pm}\rangle$] lines, three additional biexcitonic lines are
observed. These lines are due to recombination from the metastable
biexciton configurations $\rm(1e^2)(1h^12h^1)_T$ to the excited
$\rm(1e^1)(2h^1)$ exciton eigenstates. Two cross-linearly polarized
lines are due to the transitions from the
$\rm(1e^2)(1h^12h^1)_{T_0}$ biexciton configuration to the excited
bright exciton eigenstates, $\rm|X^{0}_{1,2,B\pm}\rangle$
and one, unpolarized, is due to the (almost) degenerate
transitions from the $\rm(1e^2)(1h^12h^1)_{T_{\pm 3}}$ biexciton
configurations to the excited dark exciton configurations, $\rm |X^{0}_{1,2,D\pm}\rangle$.
The observed emission intensity ratios of 1:1:4 is straightforward
to understand~\cite{Kodriano10}, as discussed above.

\begin{figure*}
\includegraphics[width=0.8\textwidth]{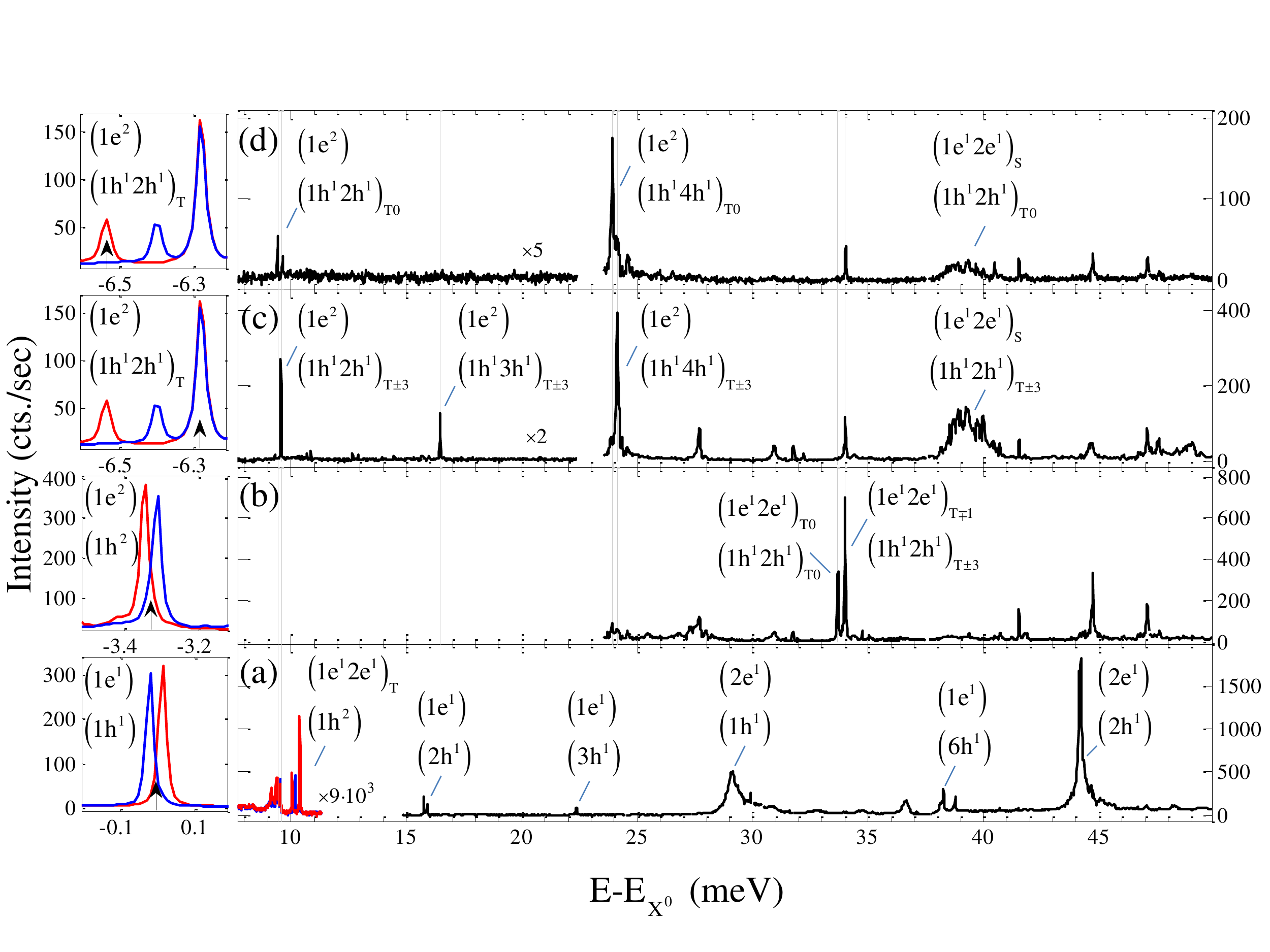}
\caption{\label{fig:PLEAll} Linearly polarized PL spectra [left
panels and the lower energy region in (a)] and PLE spectra (right
panels) of a single QD. The PLE in (a) is measured by continuous
scan of one laser's emission energy. The PLE spectra in (b)-(d) are
measured with one laser's energy tuned to the excitonic resonance at
29 meV as shown in (a), while the energy of the second laser is
continuously scanned. The PL line monitored in each case is marked
on the corresponding left panel by a vertical black arrow. The
assignment of the measured resonances is given by the state, which
the QD is excited to, above each resonance.}
\end{figure*}

In Fig.\ \ref{fig:PLEAll} we present PLE spectra of the neutral
excitonic and biexcitonic PL lines. Each panel in the figure
presents PLE spectrum of the PL spectral position marked by the
vertical arrow on the expanded scale PL spectrum to the left of the
panel. This set of measurements combined with additional
measurements (discussed below) and the intuition that we gained from
the model outlined above, allow us to resolve and identify most of
the observed one- and two- photon resonances. The identified optical
transitions are marked above the observed resonances.

\subsection{Identification of excitonic lines}
Fig.\ \ref{fig:PLEAll}(a) displays single photon absorption
resonances. When a photon is resonantly absorbed by the empty QD it
enhances the emission from the exciton lines. The spectrum is
dominated by the $\rm(2e^1)(2h^1)$ absorption resonance. This
excitonic state in which both the electron and the heavy-hole are in
their second $\rm p_H$-like orbital mode, is particularly strong due
to the large overlap between the orbitals of the two carriers. The
PLE spectrum contains additional, almost an order of magnitude
weaker sharp resonances. These resonances are due to ``non-diagonal"
excitonic states, in which the electron and the heavy-hole differ in
their orbital mode's symmetry. As a result, the spatial overlap
between their modes is small and the oscillator strength for the
optical transition is reduced. The non-diagonal transitions that we
clearly identify are the $\rm(1e^1)(6h^1)$ in
which the electron is in its first, s-like orbital mode and the
hole is in its $\rm d_{VV}$-like modes. The oscillator strength for these transitions does not
vanish, since there is some amount of overlap between the s-like and
$\rm d_{VV}$-like orbitals which are of even symmetry~\cite{Miller85}.

More surprising, is the observation of non-diagonal excitonic
transitions between orbitals of different symmetries, like the
$\rm(1e^1)(2h^1)$. This transition, which is the lowest energy
resonance in the exciton PLE spectrum, is unambiguously identified
by its spectral position and spectral shape. As expected, it is a
cross linearly polarized doublet, with the same splitting and the
same energy-order of polarizations as the PL line due to the optical
transition from the $\rm(1e^2)(2h^11h^1)_{T_{0}}$ spin blockaded
biexciton to this [$\rm(1e^1)(2h^1)$] excited non-diagonal exciton
states [Fig.\ \ref{fig:PL}]. In both cases the spectral shape is
dictated by the same final exciton states. Similarly, we identified
the next in energy order doublet as the non-diagonal transitions to
the bright levels of the $\rm(1e^1)(3h^1)$ exciton. In these
resonances [$\rm(1e^1)(2h^1)$ and $\rm(1e^1)(3h^1)$], the electron
is excited into the first, s-like, symmetric orbital mode, while the
hole is excited into the second, $\rm p_H$-like, and third, $\rm
p_V$-like, antisymmetric mode, respectively. These optical
transitions are therefore expected to be forbidden since the orbital
modes' overlap vanishes. Their appearance indicates some symmetry
breaking, possibly resulting in mixing with other
bands~\cite{Warming09,Siebert09}.

Another important mechanism which permits these symmetry
forbidden transitions is provided by phonon induced mixing. This
mixing is particularly strong when the phonon energy resonates with
the single carrier's energy levels separation~\cite{Hameau99}. Clear
evidence for such type of mixing induced excitation is seen in the
spectrally broad resonance 29 meV above the exciton line. This
energy separation characterizes the energy of LO phonons in
compounds of GaAs and
InAs~\cite{Sarkar05,Sarkar08,Lemaitre01,Findeis00}. The $\rm
In_{x}Ga_{1-x}As$ optical phonon closely resonates with the 1e-2e
energy levels separation, resulting in an enhanced absorption in
this spectral domain. This observation is also supported by the fact
that the $\rm(2e^1)(2h^1)$ resonance is higher in energy by about 29
meV from the $\rm(1e^1)(2h^1)$ resonance, as expected.

\subsection{Identification of biexcitonic lines}
In Fig.\ \ref{fig:PLEAll}(b-d) PLE spectra of the biexcitonic lines
are presented. During these measurements one laser was tuned into
the broad excitonic resonance at 29 meV, thereby populating the QD with a
bright exciton. The second laser energy was then continuously varied
while the emission from one of the biexciton lines was monitored.

The PLE spectrum of the ground biexciton doublet, $\rm(1e^2)(1h^2)
\longrightarrow  \rm(1e^1)(1h^1)$ is presented in Fig.\
\ref{fig:PLEAll}(b). The allowed transitions from the bright exciton
states (total spin $\pm 1$) into the e-triplet-h-triplet biexciton
states: $\rm(1e^12e^1)_{T_0}(1h^12h^1)_{T_0}$ (total spin zero) and
$\rm(1e^12e^1)_{T_{\pm 1}}(1h^12h^1)_{T_{\mp 3}}$ (total spin $\pm
2$) are clearly observed, dominating, as expected, this spectrum.

In Fig.\ \ref{fig:PLEAll}(c) the PLE is monitored by the PL line
which corresponds to the decay of the spin blockaded metastable
biexciton, $\rm(1e^2)(1h^12h^1)_{T_{\pm 3}}$, by recombination of a
ground e-h pair, to the excited dark exciton states, $\rm{
|X^{0}_{1,2,D \pm}\rangle}$.
This PLE spectrum is dominated by e-singlet-h-triplet resonances,
just like the resonance from which the light is monitored. The
absorption resonance transitions from the ground dark exciton $\rm{
|X^{0}_{1,1,D \pm}}\rangle$ directly to the monitored resonace
[$\rm(1e^2)(1h^12h^1)_{T_{\pm 3}}$], by photogeneration of an Oe=1
Oh=2 e-h pair, is clearly identified as the lowest energy resonance
in this PLE spectrum. Likewise, the resonances in which the hole is
excited into the Oh=3 and Oh=4 orbitals,
[$\rm(1e^2)(1h^13h^1)_{T_{\pm 3}}$ and $\rm(1e^2)(1h^14h^1)_{T_{\pm
3}}$, respectively] are clearly identified as well. Photogenerated
holes in these resonances nonradiatively relax to the Oh=2 level,
where recombination occurs, since further non-radiative relaxation
is spin blockaded~\cite{Poem10, Kodriano10}.

In addition, a broad resonance is observed $\sim$29 meV above the
$\rm(1e^2)(1h^12h^1)_{T_{\pm 3}}$ biexciton resonance. This
resonance is due to absorption into the
$\rm(1e^12e^1)_S(1h^12h^1)_{T_{\pm 3}}$. This state is strongly
coupled to the $\rm(1e^2)(1h^12h^1)_{T_{\pm 3}}$, by a one LO
phonon, in a similar way to the coupling between the $\rm(2e^1)(1h^1)$
and the $\rm(1e^1)(1h^1)$ bright exciton states [Fig.\
\ref{fig:PLEAll}(a)].

Similar spectral features are observed in Fig.\ \ref{fig:PLEAll}(d)
where the PLE is monitored through the decay of the metastable
biexciton $\rm(1e^2)(1h^12h^1)_{T_0}$ to the excited bright exciton
state $\rm |X^{0}_{1,2,B +}\rangle$. In this spectrum the absorption
resonances from the bright exciton states to the
$\rm(1e^2)(1h^12h^1)_{T_0}$ and the $\rm(1e^2)(1h^14h^1)_{T_0}$
states are identified. The weaker resonant absorption into the
$\rm(1e^2)(1h^13h^1)_{T_0}$ state, is missing from this spectrum due
to poor signal to noise ratio.

We note that the energy difference between the optical transitions
$\rm(1e^1)(1h^1)\longrightarrow (1e^2)(1h^12h^1)_{T_0}$ and $\rm (1e^2)(1h^12h^1)_{T_0}\longrightarrow(1e^1)(2h^1)$,
is 15.7 meV. As expected, this difference exactly matches the energy
of the optical transition from the vacuum into the first excited
exciton state $\rm (1e^1)(2h^1)$.

The transitions to the states $\rm(1e^2)(1h^12h^1)_{T_m}$ which are
clearly observed in Fig.\ \ref{fig:PLEAll}(c) and (d), are absent
from the PLE spectrum of the ground biexciton [Fig.\
\ref{fig:PLEAll}(b)]. This is due to the fact that in these cases
the emitting state is directly excited and no intermediate non-radiative
relaxation process is required. This is not the case when
the $\rm(1e^2)(1h^14h^1)_{T_0}$ state is excited. Here, since
non-radiative relaxation of the hole must occur prior to the
recombination, the resonance is weakly observed in the PLE spectrum
of the ground biexciton state, as well. This means that in the
relaxation process of the hole from the Oh=4 to the Oh=2 orbital
state, its spin may slightly scatter~\cite{Poem10}. Last, we note
that the resonances $\rm(1e^12e^1)_{T_{\pm 1}}(1h^12h^1)_{T_{\mp3}}$
and $\rm(1e^12e^1)_{T_0}(1h^12h^1)_{T_0}$ which are due to optical
transitions from the bright exciton states are only observed in the
PLE spectrum of the ground biexciton state [Fig.\
\ref{fig:PLEAll}(b)]. Similarly, the resonances
$\rm(1e^12e^1)_{T_{\pm 1}}(1h^12h^1)_{T_{0}}$ and
$\rm(1e^12e^1)_{T_0}(1h^12h^1)_{T_{\pm3}}$, which are due to optical
transitions from the dark exciton states are only observed in PLE
spectra of the spin blockaded biexcitons [Fig.\ \ref{fig:PLEAll}(c)
and (d)]. We note however, that the bright exciton resonances
$\rm(1e^12e^1)_{T_{\pm 1}}(1h^12h^1)_{T_{\mp3}}$ spectrally overlap
with the dark exciton resonances $\rm(1e^12e^1)_{T_{\pm
1}}(1h^12h^1)_{T_{0}}$ and
$\rm(1e^12e^1)_{T_0}(1h^12h^1)_{T_{\pm3}}$, and therefore their
final identification is also based on polarization sensitive and
time resolved spectroscopy as explained below.

\begin{figure}
\includegraphics[width=0.48\textwidth]{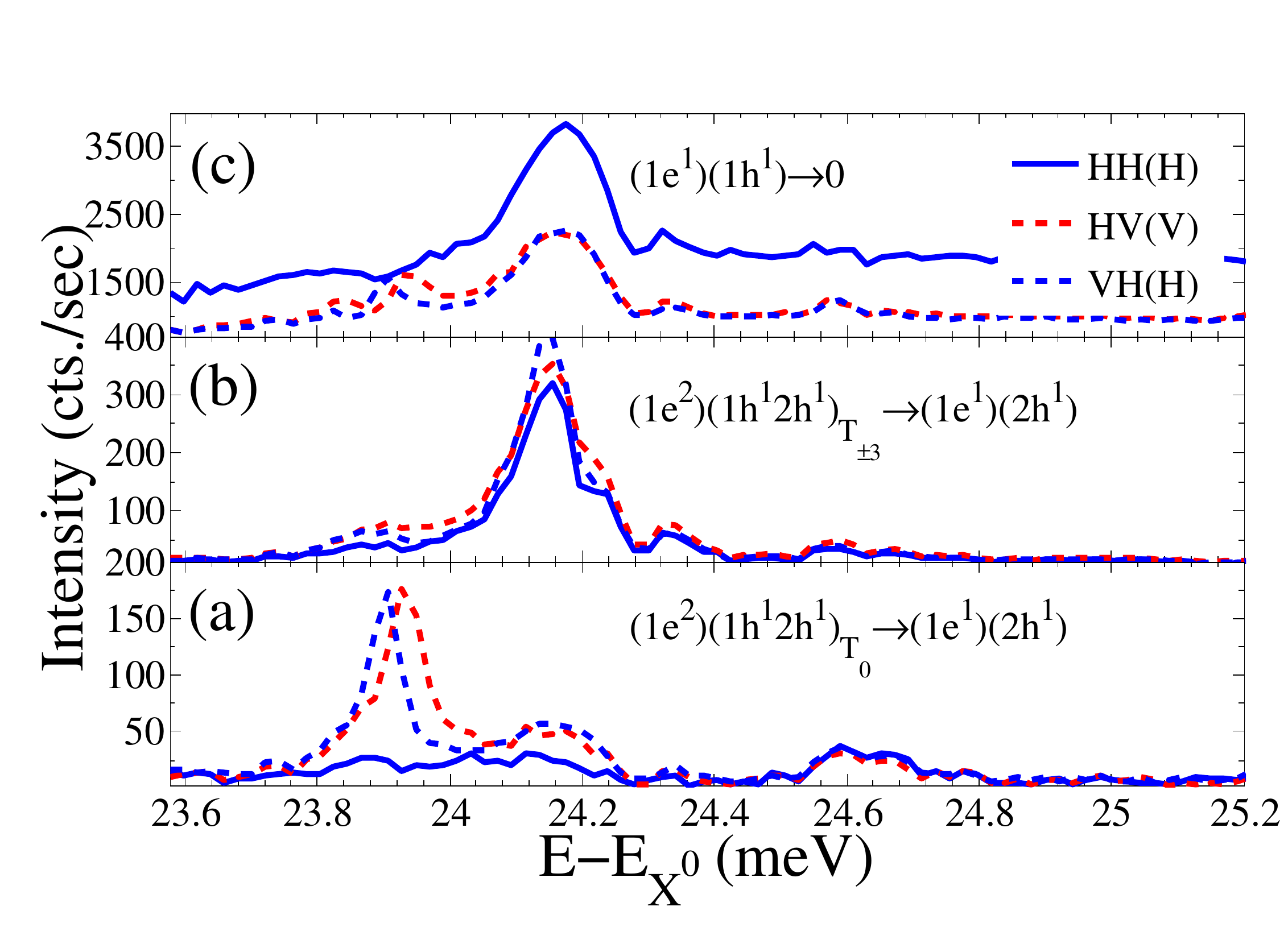}
\caption{\label{fig:Exp_SD_TT} Expanded energy scale presentation of
the two lasers polarization sensitive PLE spectra revealing the
optical transitions into the biexcitonic resonances
$\rm(1e^2)(1h^14h^1)_{T}$. The spectra are monitored by the PL lines
of the transitions from the metastable biexcitons
$\rm(1e^2)(1h^12h^1)_{T_0}$ (a) and $\rm(1e^2)(1h^12h^1)_{T_{\pm3}}$
(b) to the excited exciton $\rm(1e^1)(2h^1)$ and the PL doublet of
the transitions from the ground state exciton $\rm(1e^1)(1h^1)$ to
the vacuum (c). Various rectilinear polarizations of the exciting
lasers and the detected, PL light are used. The first letter denotes
the polarization of the laser tuned to the exciton resonance, the
second, that to the biexciton, and the third, in parentheses, that
of the detected emission. The two cross-polarized curves in the
ground state exciton (c) are multiplied by 3.}
\end{figure}

In Fig.\ \ref{fig:Exp_SD_TT}, we present examples for the use of
polarization sensitive spectroscopy as a tool for verifying the
identity of the observed spectral resonances. The PLE resonances as
monitored by the four biexcitonic PL transitions and by the ground
exciton state, are displayed in the figure for various combinations
of rectilinear polarizations of the exciting two lasers and the
detected PL. Since the figure describes e-singlet-h-triplet
resonances, the experimentally measured optical transitions and
their polarization selection rules can be directly compared with the
theoretical expectations outlined in Fig.\ \ref{fig:XX0SD}. The
characteristic three lines structure of the optical transition into
the e-singlet-h-triplet state is clearly resolved in Fig.\
\ref{fig:Exp_SD_TT}. The lowest energy biexcitonic doublet is
crossed linearly polarized, since each component is due to
excitation of a different bright exciton eigenstate. The high energy
line is unpolarized, and its intensity is twice stronger, (even in
rectilinear polarization) since it gets contributions from the two
optically allowed transitions of the dark exciton
eigenstates~\cite{Poem210}. We note in particular, that the energy
separation between the cross-linearly polarized components of the
lower energy doublet, exactly matches, as expected, that of the
bright exciton (-34 $\rm\mu eV$). The two lasers PLE spectrum of the
ground state exciton [Fig.\ \ref{fig:Exp_SD_TT}(c)] reveals a
striking difference between transitions from the bright exciton
states and transitions from the dark ones. In The first type of
transitions population from the bright exciton is transferred into
the biexciton state, in which polarization memory is totally lost,
and thus the polarized PL emission is reduced. In the latter type
population is transferred from the dark exciton into the bright
exciton state due to the rxcitation to the biexciton state.
Therefore, in this case the PL emission from the bright exciton
states is enhanced.

\begin{figure}
\includegraphics[width=0.48\textwidth]{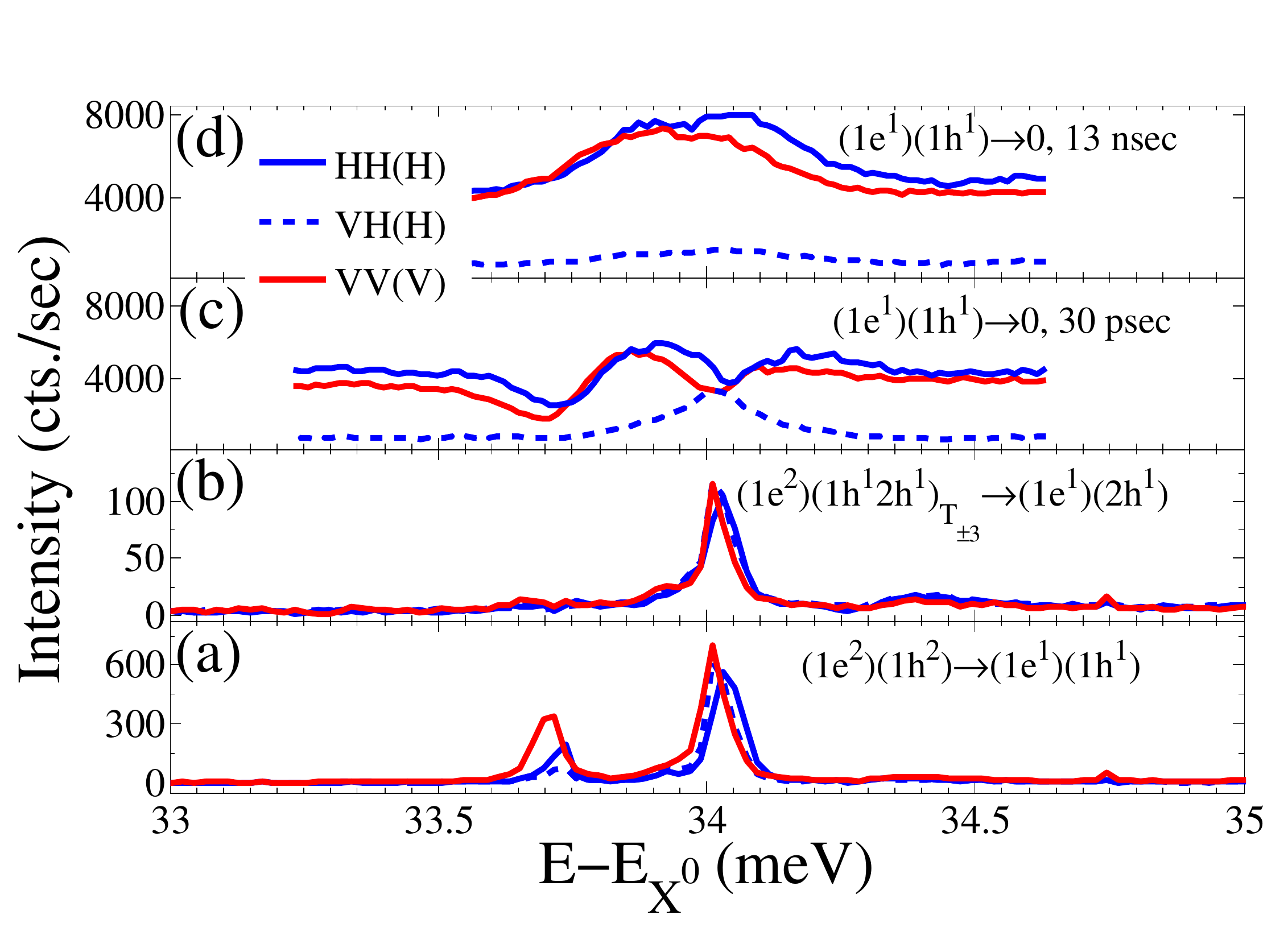}
\caption{\label{fig:Exp_SD_TT2} Expanded energy scale presentation
of the two-color, polarization sensitive PLE spectra revealing the
optical transitions into the biexcitonic resonances
$\rm(1e^12e^1)_T(1h^12h^1)_T$. The spectra are monitored by the PL
from the ground state biexciton lines (a), the metastable biexciton
$\rm(1e^2)(1h^12h^1)_{T_{\pm 3}}$ (b), and the ground state exciton
$\rm(1e^1)(1h^1)$ with two pulsed lasers at 30 psec delay (c) and 13
nsec delay (d). Various rectilinear polarizations of the exciting
lasers and the detected, PL emission are used. The first letter
denotes the polarization of the laser tuned to the exciton
resonance, the second, that to the biexciton, and the third, in
parentheses, that of the detected emission.}
\end{figure}

In Fig.\ \ref{fig:Exp_SD_TT2}, we use similar methods for studying
the richer spectrum of the e-triplet - h-triplet resonances. In this
figure the optical transitions into the
$\rm(1e^12e^1)_{T}(1h^12h^1)_{T}$ biexciton states are studied and
the experimentally measured transitions should be compared with the
theoretical considerations outlined in Fig.\ \ref{fig:XX016}. As can
be seen in Fig.\ \ref{fig:XX016}, there are six optical transitions
from the bright exciton states into the e-triplet - h-triplet
states, and four optical transitions from the dark exciton states.
The lowest energy transitions are the cross-linearly polarized
doublet due to optical transitions from the bright exciton states
into the $\rm(1e^12e^1)_{T_0}(1h^12h^1)_{T_0}$ biexciton. As
mentioned above, this biexciton resonance is only observed in the PL
from the ground biexciton states. For these optical transitions to
occur, both lasers should be co-linearly polarized, as indeed the
data show [Fig.\ \ref{fig:Exp_SD_TT2}(a)]. In addition there is a
higher energy doublet due to the four optical transitions from the
bright exciton states into the symmetric and anti-symmetric
biexciton states of total spin projection 2. Since these biexciton
states are almost degenerate (in a similar way to the dark exciton
states), the four transitions form an unpolarized doublet which is
twice as strong as the lower energy cross-polarized one. Again, this
is exactly what one sees in the polarization sensitive PLE spectrum
of the ground biexciton [Fig.\ \ref{fig:Exp_SD_TT2}(a)].

In Fig.\ \ref{fig:Exp_SD_TT2}(b) a strong resonance in the PLE
spectrum of the PL line $\rm(1e^2)(1h^12h^1)_{T_{\pm3}}$ to
$\rm(1e^1)_{\pm1/2}(2h^1)_{\pm3/2}$ is observed. The resonances in
this spectrum are expected to result mainly from excitations of the
dark exciton states. As seen in see Fig.\ \ref{fig:XX016}, the
optical transitions from the dark exciton states are expected to
form a cross linearly polarized doublet. Unfortunately, this doublet
spectrally overlaps the unpolarized doublet due to transitions from
the bright exciton states. We use time resolved pulsed PLE
spectroscopy in order to resolve these transitions.

In Fig.\ \ref{fig:Exp_SD_TT2}(c) we show two-pulse polarization
sensitive PLE spectra of the bright exciton lines when the temporal
separation between the two pulses is relatively short (30 psec),
while in Fig.\ \ref{fig:Exp_SD_TT2}(d) these spectra are shown for
the case in which the temporal separation is 13 nsec. While in the
first case, immediately after the photogeneration the exciton
population is bright, in the second case only dark exciton
population lasts. The PLE spectroscopy reveals this fact in the
following way: When the second pulse is tuned into a bright exciton
to biexciton transition, the PL signal from the exciton lines is
reduced, since from the biexciton state part of the population does
not return to the monitored exciton state. This is particularly true
for co-polarized pulses, since the polarization memory is lost in
the biexciton states. Therefore, bright exciton transitions are seen
as dips in the PLE spectrum of the exciton for co-polarized pulses
and as peaks for cross-polarized pulses. Dark exciton transitions
are always obtained as peaks in the PLE spectrum of the exciton,
since they transfer dark population into bright one through the
biexciton states. Thus, the polarized nature of the optical
transitions from the dark exciton states into the $J= \pm 1; \pm 3$
biexciton states are clearly revealed in the polarization sensitive
PLE spectra of the exciton in Fig.\ \ref{fig:Exp_SD_TT2}(d).

\begin{figure*}
\includegraphics[width=0.85\textwidth]{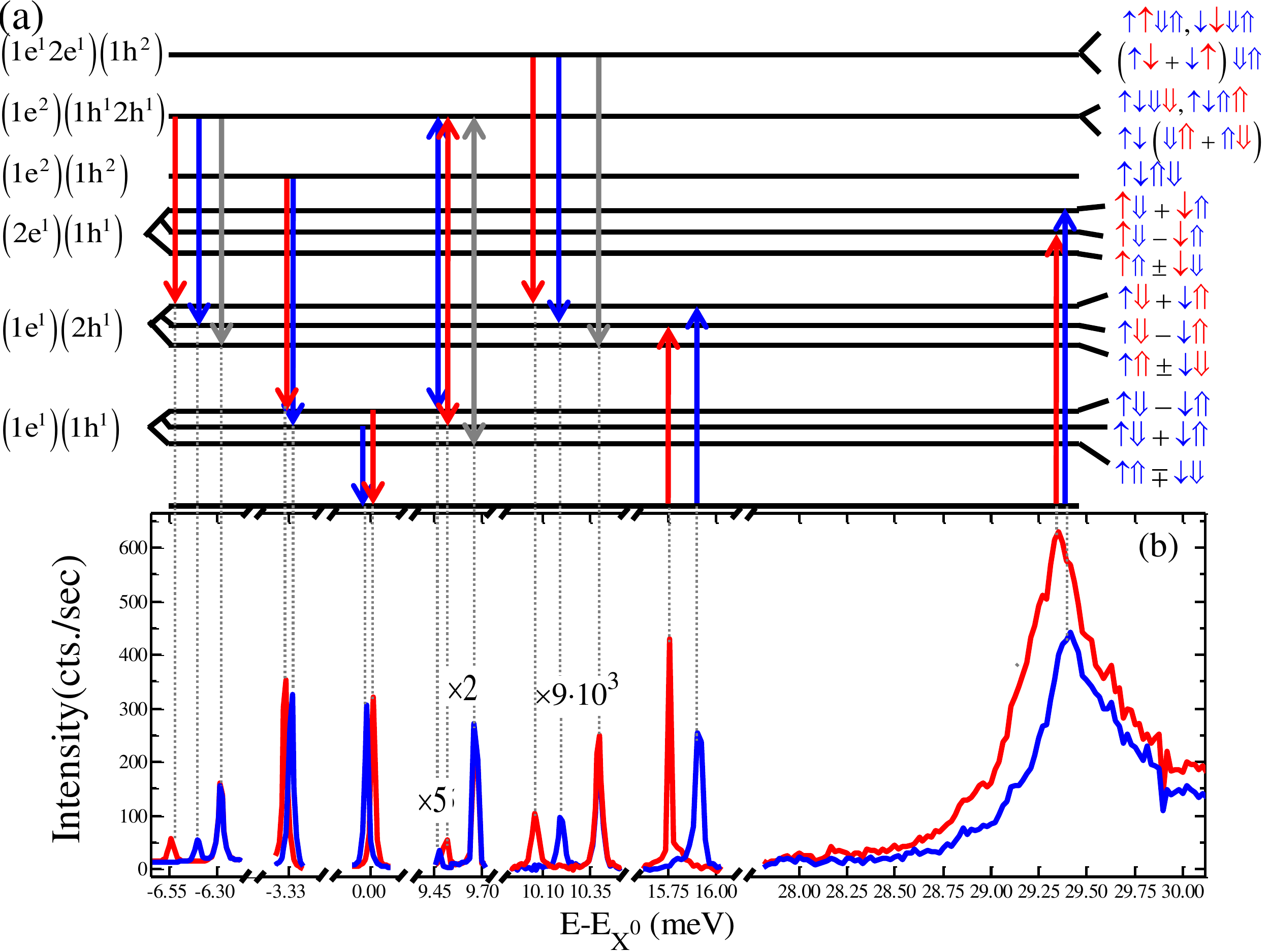}
\caption{\label{fig:Exp_1Se2Th_2Te1Sh} (a) Schematic description of
the excitonic and biexcitonic energy levels and carriers' spin
wavefunctions associated with the first and the second orbital
modes. The optical transitions between these levels and in
particular the `non-diagonal' ones are represented by vertical
arrows. Blue (red) downward (upward) arrow describes horizontally
(vertically) linearly polarized emission (absorption) transition.
Gray arrows represent unpolarized transitions. The optical
transitions in (a) are linked with the experimentally measured
polarization sensitive PL and PLE spectra in (b) and in (c),
respectively. Solid blue (red) line represent horizontal (vertical)
polarization.}
\end{figure*}
\begin{figure}
\includegraphics[width=0.43\textwidth]{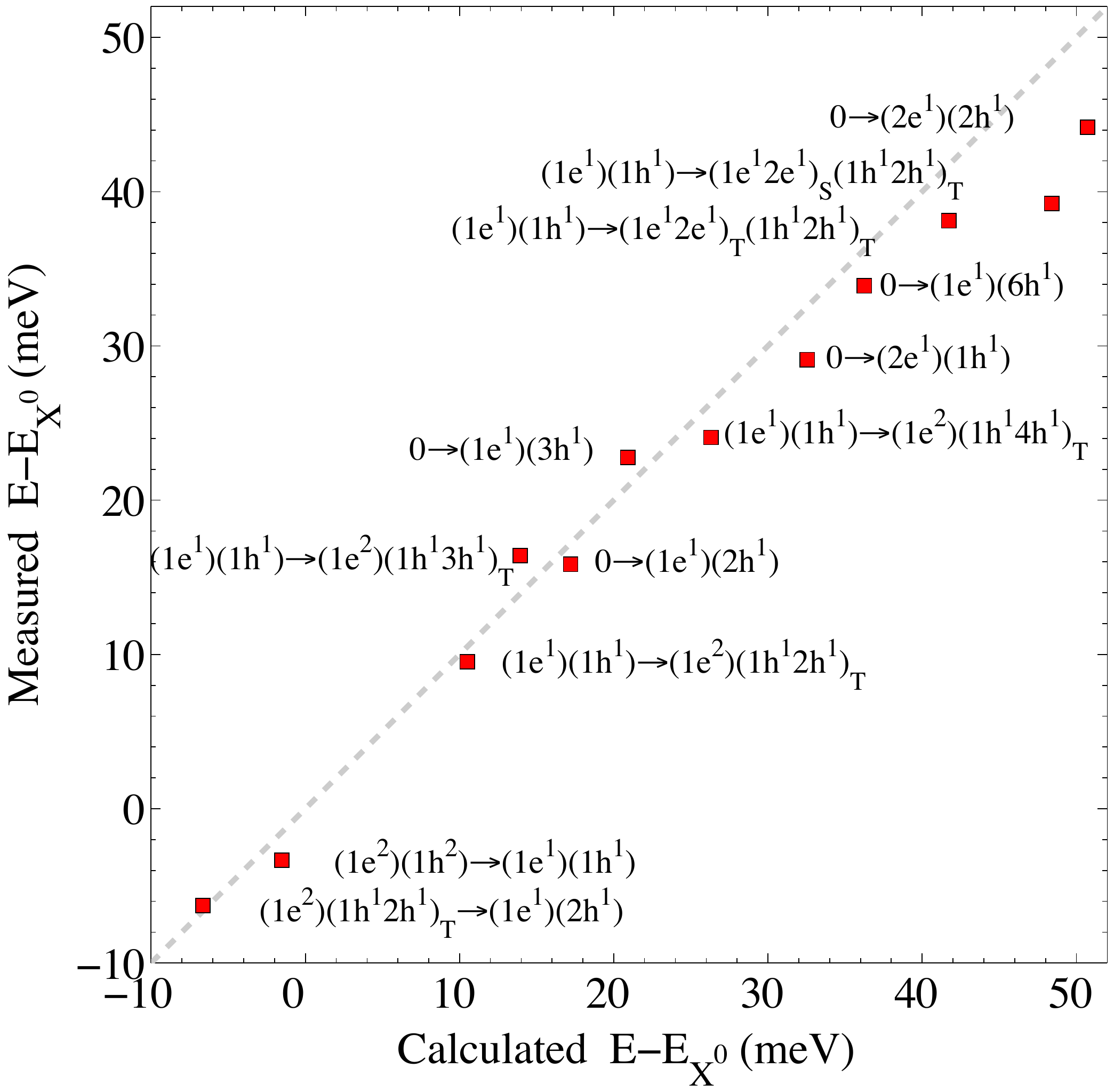}
\caption{\label{fig:COMPARISON} Comparison between the measured and
calculated excitonic and biexcitonic transitions.}
\end{figure}
\subsection{The non-diagonal optical transitions}
In Fig.\ \ref{fig:Exp_1Se2Th_2Te1Sh} we focus our attention on the
``non-diagonal" optical transitions that we identified in the PL and
the one and two color PLE spectra. Only transitions which include
 $p_H$-like orbitals are considered. Fig.\
\ref{fig:Exp_1Se2Th_2Te1Sh}(a) schematically describes the relevant
excitonic and biexcitonic energy levels and the optical transitions
between them, once the non-diagonal transitions  become allowed. The
(un-normalized) spin wavefunctions are described to the left of each
level. Downward (upward) vertical arrow describes emission
(absorption) and blue (red) stands for H (V) polarization. Gray
arrows describe unpolarized transitions. We note that the
transitions between the $\rm(1e^2)(1h^12h^1)_T$ biexciton states and
the $\rm(1e^1)(1h^1)$ exciton states are observed both in PL and in
PLE spectra.

The spectra are characterized by two repeating patterns: The first
one is a cross-linearly polarized doublet. This doublet is due to
the anisotropic e-h exchange induced splitting of the bright exciton
states. In these doublets the symmetric state is lower (higher) in energy than the antisymmetric state
for diagonal (non-diagonal) optical transitions.
The second pattern has three spectral lines: a higher energy
unpolarized line and a lower energy cross-linearly polarized
doublet. This pattern is due to transitions from
exciton to biexciton singlet-triplet states. The doublet is due to
transitions from the bright exciton states and the unpolarized line
is due to transitions from the dark exciton states.
The optical transitions which are schematically described in Fig.\
\ref{fig:Exp_1Se2Th_2Te1Sh}(a) are linked to the experimentally
measured transitions in the PL and in the PLE spectra [Fig.\ \ref{fig:Exp_1Se2Th_2Te1Sh}(b)].

\begin{table}
\caption{\label{tb:par} Spectroscopically extracted e-h exchange
interaction energies. }
\begin{ruledtabular}
\begin{tabular}{cc}
Energy separation &Measured value [$\rm\mu eV$]\\
\hline
$\Delta^{1,1}_0$&123\\
$\Delta^{1,1}_1$&-34\\
$\Delta^{1,2}_0$&200\\
$\Delta^{1,2}_1$&151\\
$\Delta^{2,1}_1$&60\\
$\Delta^{2,2}_1$\footnote{extracted from the $\rm(2e^1)(2h^1)$
doublet and from the e-triplet-h-triplet
resonances. (see Fig.\ \ref{fig:XX016}).}&60\\
\end{tabular}
\end{ruledtabular}
\end{table}

Our ability to unambiguously identify all these non-diagonal optical
transitions allows us to fully characterize the QD in terms of
single carriers' orbital mode energies and various interaction terms
between carrier pairs. The energies extracted from our spectroscopy
are summarized in Table \ref{tb:par}.

In Fig.\ \ref{fig:COMPARISON} we compare between the measured
excitonic and biexcitonic optical transitions and the calculated
ones. Good agreement is achieved using the QD parameters listed in
Table~\ref{tb:par2}. One notes that transitions which include the
second single electron state 2e, deviate the most from the
calculated ones. We believe that it is due to the optical phonon
induced coupling between this state and the first electronic state.
Our model does not consider this coupling.

Finally, we note that the three optical transitions from the
$\rm(1e^12e^1)_{T}(1h^2)$ to the $\rm(1e^1)(2h^1)$ states are
observed in the measured PL spectrum. This confirms our many-body description, as discussed in
section~\ref{sec:TheoManyBody}.

\begin{table}
\caption{\label{tb:par2} The parameters used for the model.}
\begin{ruledtabular}
\begin{tabular}{ll}
Parameter &Value\\
\hline
$M_{\bot,h}^{*}$\footnote[1]{Reference~\cite{Poem07}.}&0.25$m_0$\\
$M_e^{*}$\footnotemark[1]&0.065$m_0$\\
$l_h^x$&53{\AA}\\
$l_e^x$&74{\AA}\\
$\xi=\frac{l_e^y}{l_e^x}=\frac{l_h^y}{l_h^x}$&0.87
\end{tabular}
\end{ruledtabular}
\end{table}

\section{\label{sec:Summary}Summary}
In summary, we presented a comprehensive study of single, neutral
semiconductor quantum dots subject to excitation by two variably
polarized resonant excitations, one to exciton resonances, and the
other to biexciton resonances. By monitoring the emission intensity
from various exciton and biexciton lines we completely characterize
the rich one- and two-photon absorption spectra of single
semiconductor quantum dots. The measured data is compared with a
many carrier theoretical model, based on simple, one band parabolic
potentials for electrons and heavy-holes. While the model provides
full understanding of the observed resonances, in terms of line
shapes, energies and polarization selection rules, it is short of
quantitatively describing intensities of various ``non-diagonal"
optical transitions and spectral features which involve strong
coupling with optical phonons. We believe that the understanding
that our study provides, should be very useful in applying
semiconductor quantum dots as devices for quantum logical gates.

\begin{acknowledgments}
The support of the US-Israel binational science foundation (BSF),
the Israeli science foundation (ISF), the ministry of science and
technology (MOST), Eranet Nano Science Consortium and that of the
Technion's RBNI are gratefully acknowledged. We also acknowledge
helpful discussions with Dr.\ Garnett Bryant.
\end{acknowledgments}

\clearpage

\end{document}